\documentclass[aps,pre,twocolumn,superscriptaddress,showpacs,showkeys]{revtex4-1}
\usepackage{graphicx}                   % for insert file eps
\usepackage{hyperref}                   % for hyper refrences
\usepackage{amsmath,amssymb}            % for insert extra math
\usepackage{epstopdf}
\usepackage{subcaption}					% for sub-figures
\usepackage{float}
\usepackage{color}
\definecolor{orange}{rgb}{1,0.5,0}
\definecolor{brown}{rgb}{0.65, 0.16, 0.16}
\definecolor{phlox}{rgb}{0.87, 0.0, 1.0}

\graphicspath{{figs/}}					% the path of image folder 

\begin{document}

	\title{ Two temperature Ising Model }

	\author{J. Cheraghalizadeh*}
	\affiliation{Department of Physics, University of Mohaghegh Ardabili, P.O. Box 179, Ardabil, Iran}
	\email{jafarcheraghalizadeh@gmail.com}
	
	\author{M. Seifi}
	\affiliation{Department of Physics, University of Mohaghegh Ardabili, P.O. Box 179, Ardabil, Iran}
	\email{maryam.seifi.physics@gmail.com}
	
	\author{Z. Ebadi}
	\affiliation{Department of Physics, University of Mohaghegh Ardabili, P.O. Box 179, Ardabil, Iran}
	\email{zahraa.ebadi@gmail.com}
	
	\author{H. Mohammadzadeh}
	\affiliation{Department of Physics, University of Mohaghegh Ardabili, P.O. Box 179, Ardabil, Iran}
	\email{h.mohammadzadeh@gmail.com}
	
	\author{M. N. Najafi}
	\affiliation{Department of Physics, University of Mohaghegh Ardabili, P.O. Box 179, Ardabil, Iran}
	\email{morteza.nattagh@gmail.com}

	\begin{abstract}
We introduce a two-temperature Ising model as a prototype of superstatistic critical phenomena. The model is described by two temperatures ($T_1,T_2$) in zero magnetic field. To predict the phase diagram and numerically estimate the exponents, we develop Metropolis and Swendsen-Wang Monte Carlo method. We observe that there is a non-trivial critical line, separating ordered and disordered phases. We propose an analytic equation for the critical line in the phase diagram. Our numerical estimation of the critical exponents illustrates that all points on the critical line belong to the ordinary Ising universality class.
	\end{abstract}

	
	\maketitle

	\section{Introduction}

%----------------------------------------------------------------------------------------
What is the impact of the fluctuations in macroscopic parameters like the temperature (as a long-standing problem) in the physical systems, and especially in the thermodynamics and statistical physics? Although the fluctuations in temperature brings the system to out-of-equilibrium phenomena, some concepts of equilibrium thermodynamics can be employed for studying them. Superstatistics as a systematic way of handling such fluctuations is an example. The superstatics cannot be defined uniquely, in general terms, however, it is called the statistics of the statistics of ordinary Boltzmann factor $\exp(-\beta E)$ ($\beta$ and $E$ being the usual inverse temperature and the energy), whose parameters have relations with the fluctuations of thermodynamic quantities such as temperature~\cite{beck2003superstatistics1}. In other words, superstatistics is a systematic way of handling the statistical systems with superpositions of various Boltzmann distributions~\cite{BECK2003267}. It was propounded by Beck and Cohen~\cite{BECK2003267,beck2004} with a primitive goal of modeling the non-Maxwell-Boltzmann statistical distributions in  out-of-equilibrium complex systems. Actually the Tsallis statistics and also Levy distributions were earlier examples of superstatistics~\cite{PhysRevLett.84.2770}, for which it was shown that the nonextensivity is given by the fluctuations of the parameters of the usual exponential distributions. The microscopic fluctuations (in the random friction forces) also was shown to lead the system with ordinary statistical mechanics to behave effectively according to more general nonextensive case. This might serve as the mechanism behind the fact that many physical systems with fluctuating temperature or energy dissipation rate are described by Tsallis statsitics~\cite{PhysRevLett.87.180601}. 
The superstatistics has been employed in many physical systems such as nuclear physics~\cite{PhysRevLett.84.2770}, turbulent fluids~\cite{PhysRevLett.91.084503,PhysRevLett.98.064502,PhysRevLett.87.180601}, solar flares~\cite{PhysRevLett.96.051103}, ultra cold gases~\cite{PhysRevLett.118.143401,ALVES2016195} and quantum entanglement~\cite{OURABAH20172659,PhysRevE.95.042111}.\\
The fluctuations in macroscopic parameters are assumed to be on a long time scale so that the system can temporarily reach local equilibrium~\cite{beck2004superstatistics}. A common approach in these systems is to equating the conditional probabilities to the probability measures of a system without fluctuation.\\
Despite of a huge literature in the application of superstatistics in non-equilibrium systems~\cite{beck2009recent,iliopoulos2019superstatistics,BRIGGS2007498}, very little attention has been paid to superstatistic critical phenomena, i.e. superstatistics in statistical systems in the vicinity of critical point. More explicitly what is the effect of temperature fluctuations in statistical models, especially in the vicinity of the critical points? As an start to this research channel, which we call superstatistic critical phenomena (SCP), we consider two-dimensional Ising model with two temperatures $(T_1,T_2)$ with zero magnetic field. Our motivation for choosing the Ising model is that it is extensively used in statistical mechanics as a prototype of equilibrium system which undergoes non-trivial phase transition, the main one being order-disorder transition at a critical temperature $T_c$ in the absence of magnetic field. Many interesting aspects of this model is known~\cite{mccoy2014two}. This involves the coexistence of percolation and magnetic phase transition~\cite{hu1984percolation}, elastic backbone transition~\cite{najafi2019elastic}, equivalence to Schramm-Loewner evolution with  $\kappa=2$~\cite{najafi2015observation}, probability measure of the order parameter~\cite{tsypin2000probability}, its relation to the free fermionic model~\cite{mussardo2010statistical,francesco2012conformal}. Also the Ising model has vastly been used as a partner on other combined statistical models. The example is the self-organization critically (SOC) on Ising percolation lattices~\cite{PhysRevLett.59.381,PhysRevA.38.364,najafi2020geometry,cheraghalizadeh2017mapping,najafi2018coupling}. The Bak-Tang-Wiesenfeld (BTW) model, has been used to investigate the movement pattern of fluid in correlated porous media~\cite{najafi2016water,PhysRevE.96.052127,PhysRevE.101.032116}, where the correlations in the porous media is modeled controlled by the Ising coupling constant and the artificial temperature. Also it is used to investigate the effect of correlated environmental disorder on critical behaviors of systems like loop-erased random walk (LERW) and self-avoiding walk (SAW) that it explain how polymers grow on correlated prose media~\cite{cheraghalizadeh2019correlation,PhysRevE.97.042128}. In light of the known properties of the Ising model, especially in the vicinity of the critical point, we are able to recover many aspects of the SCP.\\

Here we consider a two-dimensional Ising model with fluctuating temperature described by a binary distribution. We develop Metropolis and also the Swendsen-Wang(SW) Monte-Carlo method~\cite{swendsen1987nonuniversal} for investigating the system numerically and analytically. We show that an order-disorder phase transition takes place over an extended line. By extracting various exponents, we illustrate that the universality class of all points on the critical line is consistent with the ordinary Ising universality class, indeed we see critical line with ordinary Ising universality class.\\

The paper is organized as follows: In the next section we introduce the problem with binary distribution. The model is introduced in this section, along with the Metropolis and SW algorithms. The numerical results are presented in Sec.~\ref{Sec:num}, were we explore the properties of the extended critical line. We close the paper by a conclusion.

%----------------------------------------------------------------------------------------
\section{THE CONSTRUCTION OF THE PROBLEM}
The stationary probability density of an equilibrium system is described by the Boltzmann factors $\exp(-\beta E)$, where $\beta$ is the inverse of temperature and $E$ is the system energy. For out-of-equilibrium systems this law is replaced by other more sophisticated scenarios, ranging from Einstein's relation of fluctuations~\cite{beck2003superstatistics1} to local equilibrium systems with temperature changing from place to place~\cite{beck2004superstatistics}. For the latter case an averaged Boltzmann factor is defined as follows

\begin{equation}
B(E) = \int_0^\infty \! f(\beta) e^{-\beta E} \, \mathrm{d}\beta,
\label{bf}
\end{equation}
where  $f(\beta)$  is superstatistical kernel, and $E$  is the total energy of the system in the respective microstate. Indeed it is the probability distribution of $\beta$ which reads:
\begin{equation}
p(E) = \frac{1}{Z}B(E),
\end{equation}
where,
\begin{equation}
Z  = \int_0^\infty \! B(E)  \, \mathrm{d}E,
\end{equation}
The kernel is positive and normalized, i.e. $\int_0^\infty \! f(\beta)\, \mathrm{d}\beta =1$. For a fixed non-fluctuating temperature $\frac{1}{\beta_0}$, the kernel is $f(\beta)=\delta(\beta - \beta_0) $ and consequently  $B(E)$ is an ordinary Boltzmann factor, where $\delta$ is Dirac delta function. Various supserstatistical kernels have been investigated in \cite{BECK2003267,SATTIN20182551,Generalized}.\\

The simplest generalization of Boltzmann factor is a system with two temperature, i.e. the system which fluctuates between two different discrete values of the temperature $\beta_1 = \frac{1}{T_1}$ and $\beta_2 = \frac{1}{T_2}$ with a same probability. The probability distribution of $\beta$ is given by
\begin{equation}
f(\beta)=\frac{1}{2}\left[ \delta (\beta - \beta_1)+\delta (\beta - \beta_2)\right].
\label{q0000}
\end{equation}
The generalization of the above distribution to $n$ temperatures is straighforward. The important question here is how the properties of the model in hand is changed under this generalization. Using Eq.(\ref{bf}), a generalized Boltzmann factor is obtained which we call two-level Boltzmann factor (2LBF) as follows
\begin{equation}
B(E) = \frac{1}{2}(e^{-\beta_1 E}+e^{-\beta_2 E}).
\label{q00}
\end{equation}
In the remaining of the paper we focus on the application of 2LBF on two-dimensional Ising model on a square lattice. The Ising Hamiltonian is defined as:
\begin{equation}
H =-J \sum_{\langle i,j \rangle}^{} \sigma_i \sigma_j - h \sum_{i}^{} \sigma_i,
\end{equation}
where $J$ is the coupling constant, $h$ is the magnetic field (which is set to zero in this paper), $\sigma_i$ and $\sigma_j$ are the spins (with values $\pm 1$) at sites $i$ and $j$ respectively, and $\langle i, j \rangle$ shows that the sites $i$ and $j$ are neighbors.\\

\begin{figure*}[t]
	\begin{subfigure}[b]{0.24\linewidth}
		\includegraphics[width=\linewidth]{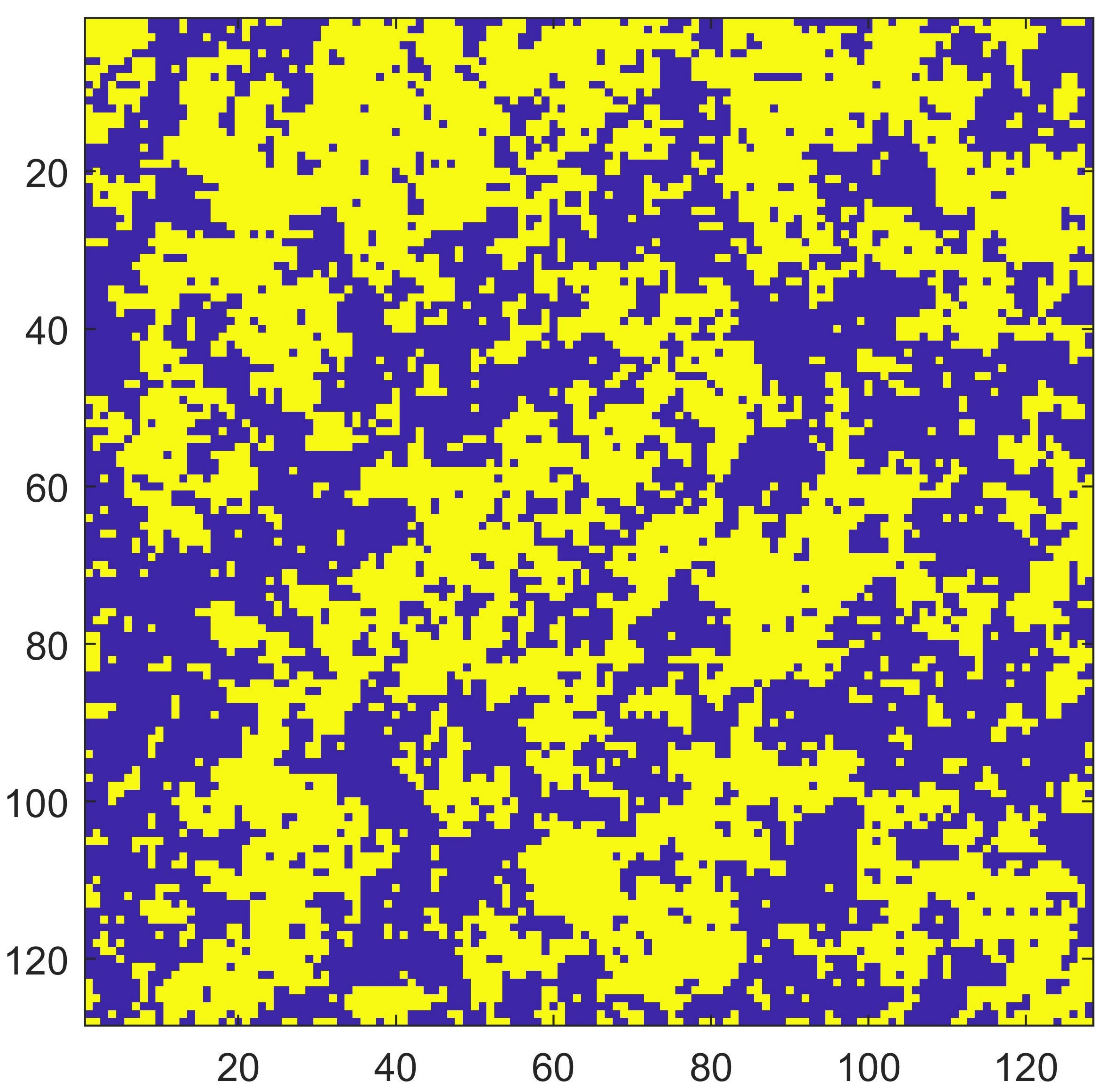}
		\caption{$T_2 = 2.50$ , $T_1 = 0.001$}
		\label{fig:t5}
	\end{subfigure}
	\begin{subfigure}[b]{0.24\linewidth}
		\includegraphics[width=\linewidth]{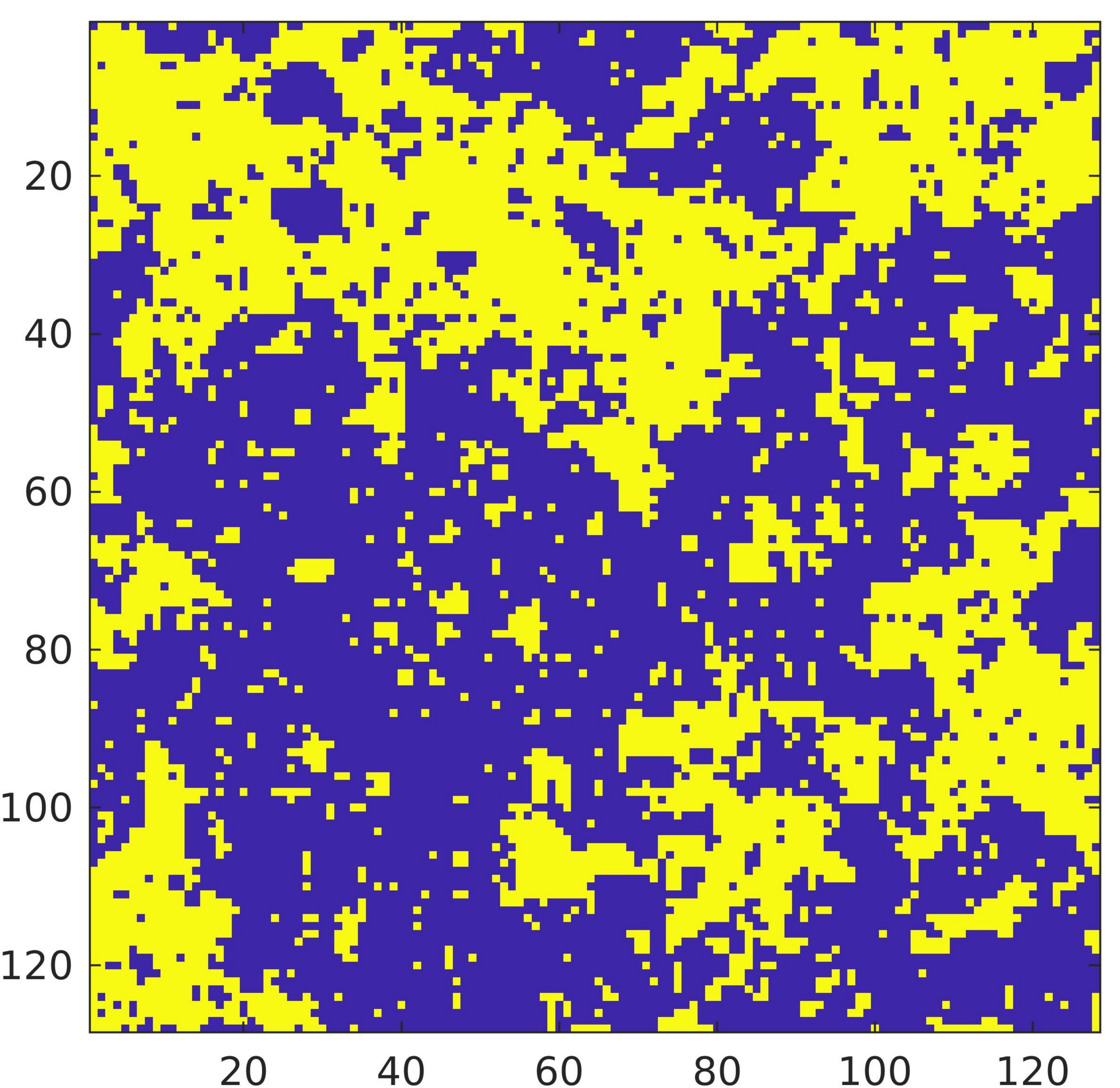}
		\caption{$T_2 = 2.50$ , $T_1 \approx T_1^{c_1}$}
		\label{fig:t6}
	\end{subfigure}
	\begin{subfigure}[b]{0.24\linewidth}
		\includegraphics[width=\linewidth]{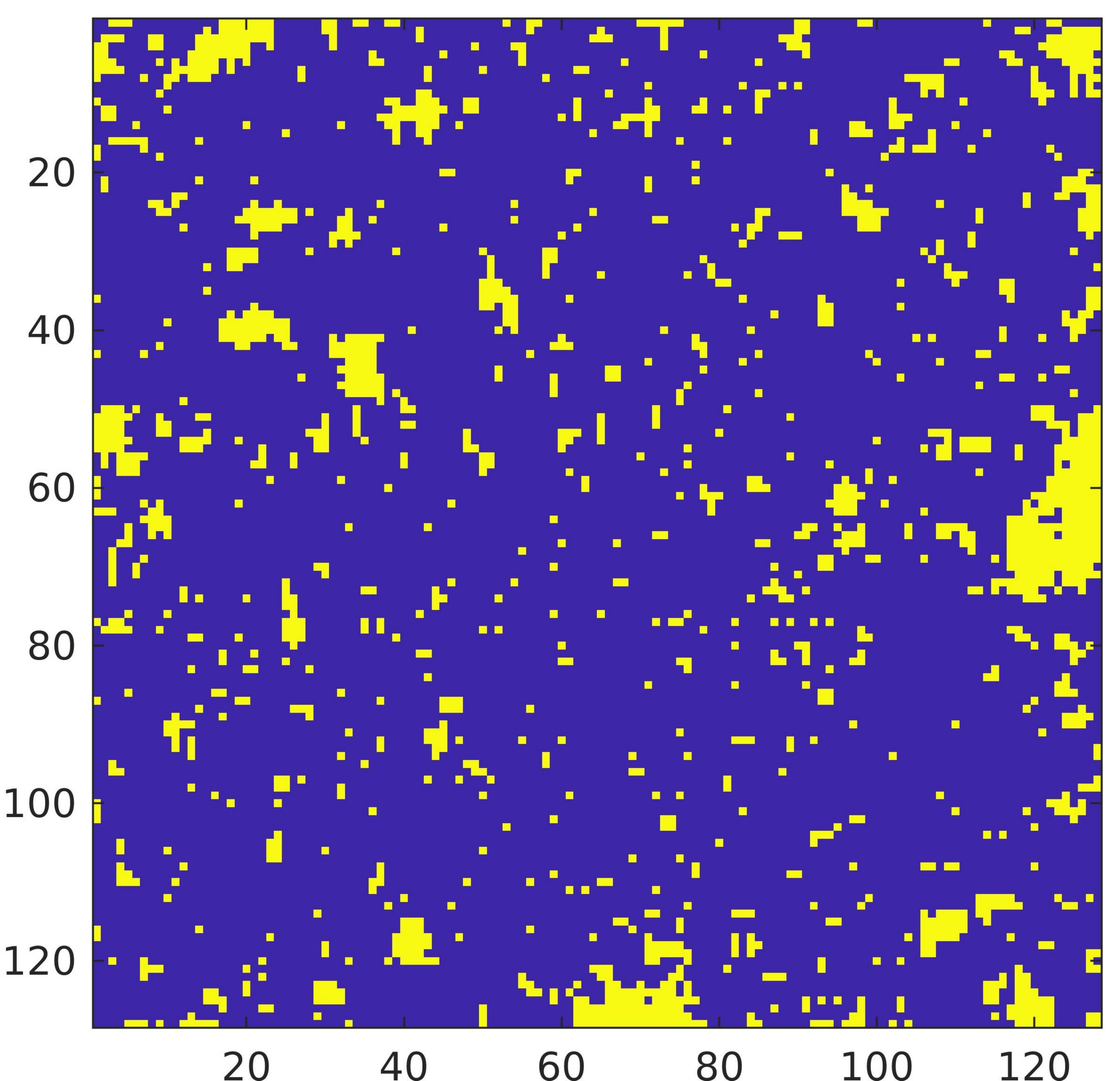}
		\caption{$T_2 = 2.50$ , $T_1 = 1.50$}
		\label{fig:t7}
	\end{subfigure}
	\begin{subfigure}[b]{0.24\linewidth}
		\includegraphics[width=\linewidth]{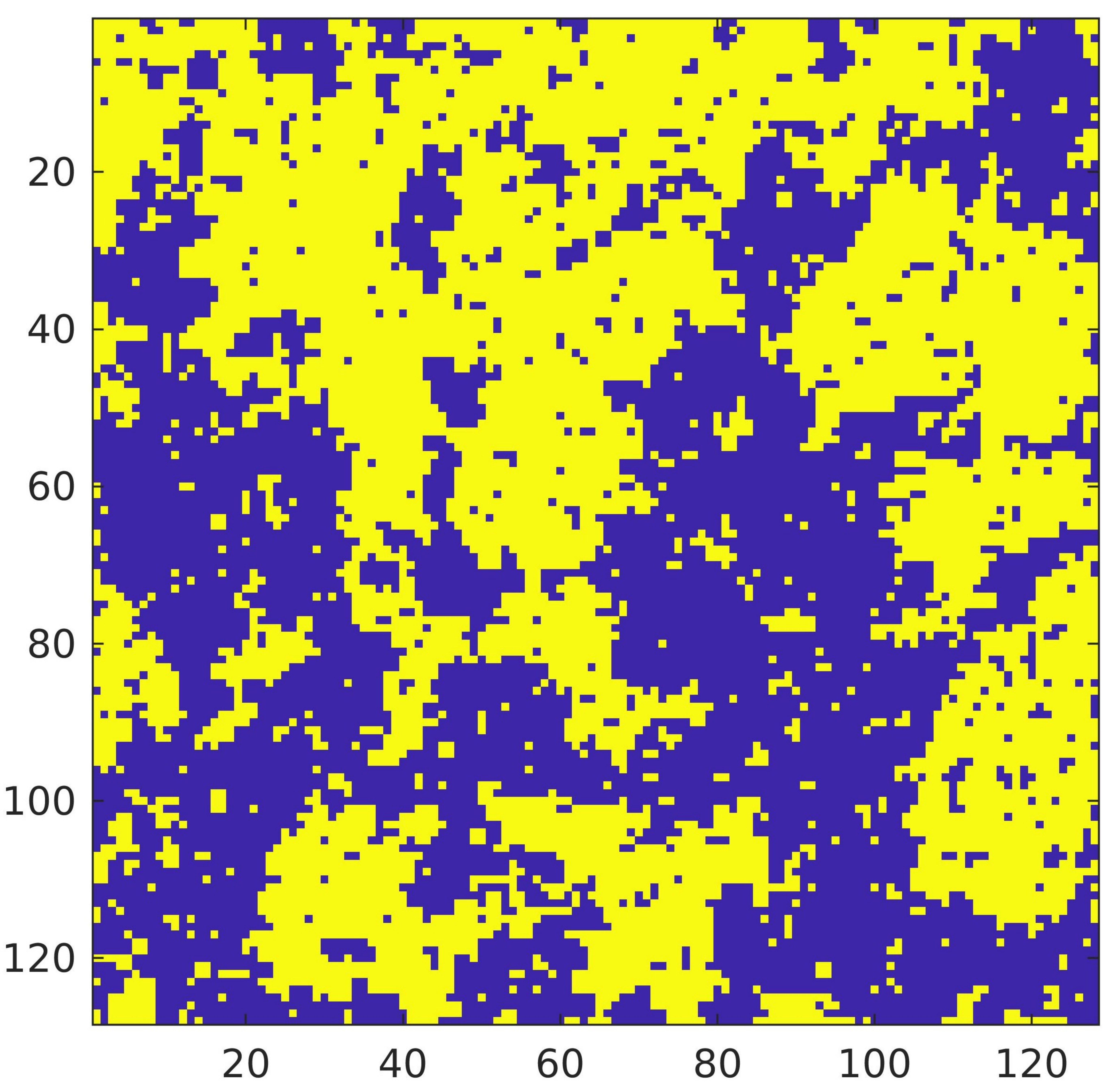}
		\caption{$T_2 = 2.50$ , $T_1  \approx T_1^{c_2}$}
		\label{fig:t8}
	\end{subfigure}
	\centering
	\caption{(Color online):The Ising samples by Metropolis steps in $128 \times 128 $ lattice size  at fixed  $T_{2} = 2.50$ for four variant temperatures $(T_1)$. For all fixed temperatures between regular Ising critical temperature ( $T_c \approx 2.26918$) and $T_D^{\models}$ there are two critical point. Yellow sites show spin up and blue shows spin down}
	\label{t56}	
\end{figure*}

For investigating the Ising model with 2LBF, we first develop Metropolis Monte Carlo schemes to investigate the problem numerically. Under a single spin flip in the Ising model the total energy is changed to $E^{\prime}=E+\delta E $, where $\delta E$ is the energy excess gained by the flip. Then according to the Metropolis method for a single temperature system, the probability of accepting this operation is
\begin{equation}
p^{\text{single temperature}}(\beta_0)=\frac{e^{-\beta_0 E'}}{e^{-\beta_0 E}} = e^{-\beta_0 \delta E },
\label{q000}
\end{equation}
To generalize this for two temperature Ising model using the generalized Boltzmann factor in Eq.(\ref{q00}), we act the same as above, this time for 2LBFs ($B(E)$) 
\begin{equation}
\begin{split}
p^{2\text{LBF}}(\beta_1,\beta_2)&\equiv\frac{B(E')}{B(E)}\\
&=\frac{e^{-\beta_1 \delta E}}{1+e^{-(\beta_1-\beta_2)E}}+\frac{e^{-\beta_2 \delta E}}{1+e^{(\beta_1-\beta_2)E}}
\end{split}
\label{q10}
\end{equation}
which reduces to readily $p^{\text{single temperature}}(\beta_0)$ in the limit $\beta_1=\beta_2=\beta_0$. For simulations, one starts from a random spin configuration and in each step choose a site randomly and apply the flip with the probability
 \begin{equation}
 p = \min [1,p^{2\text{LBF}}(\beta_1,\beta_2)].
 \label{a1}
 \end{equation}
For each temperature couple ($T_1,T_2$), this process continues until reaching stationary state in the energy landscape. Some example has been shown in Fig.(\ref{t56}).\\

\begin{figure*}[t]
	\begin{subfigure}[b]{0.24\linewidth}
		\includegraphics[width=\linewidth]{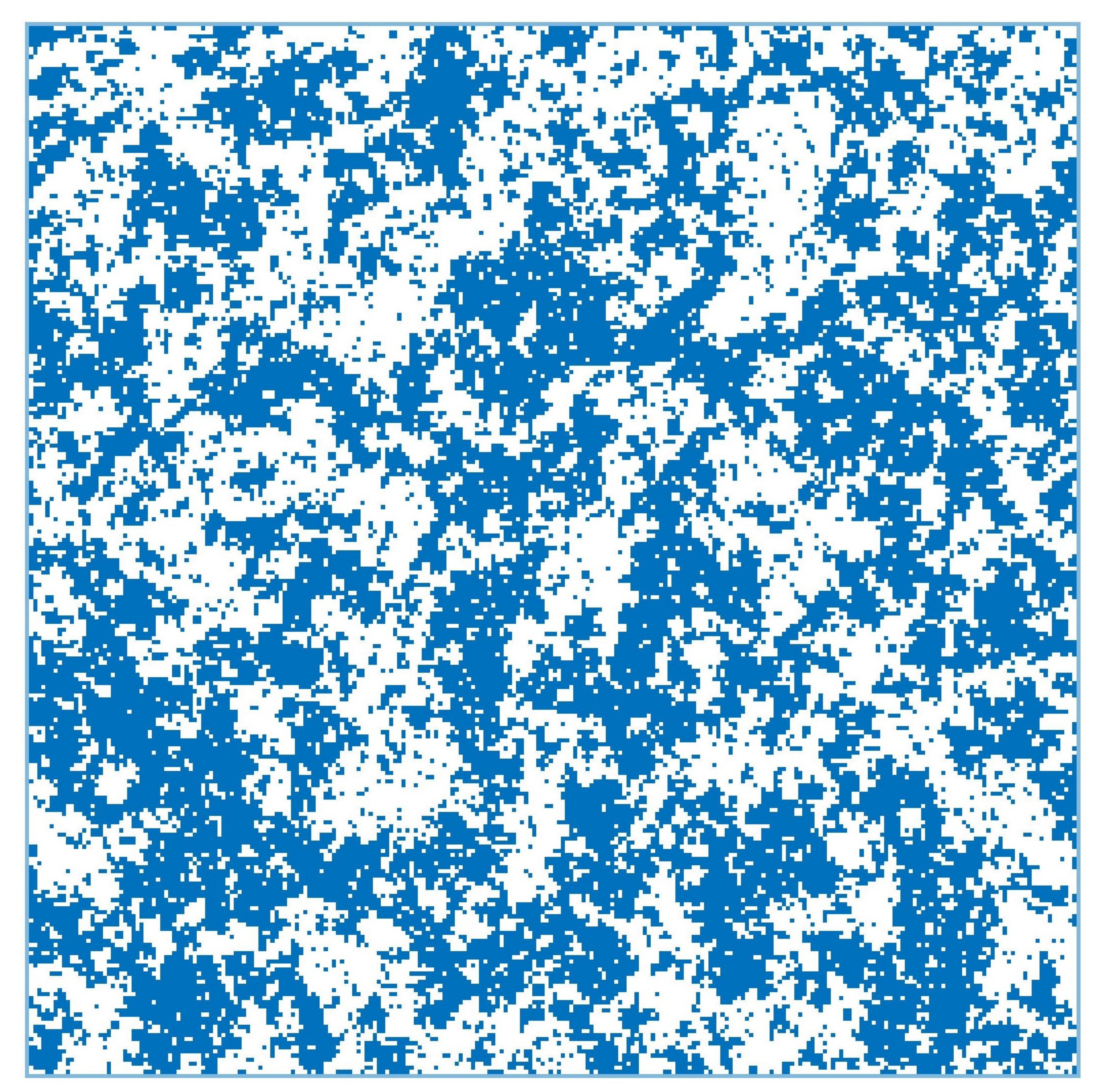}
		\caption{$T_{\text{2}} = 2.50$ , $T_1 = 0.001$}
		\label{fig:a}
	\end{subfigure}
	\begin{subfigure}[b]{0.24\linewidth}
		\includegraphics[width=\linewidth]{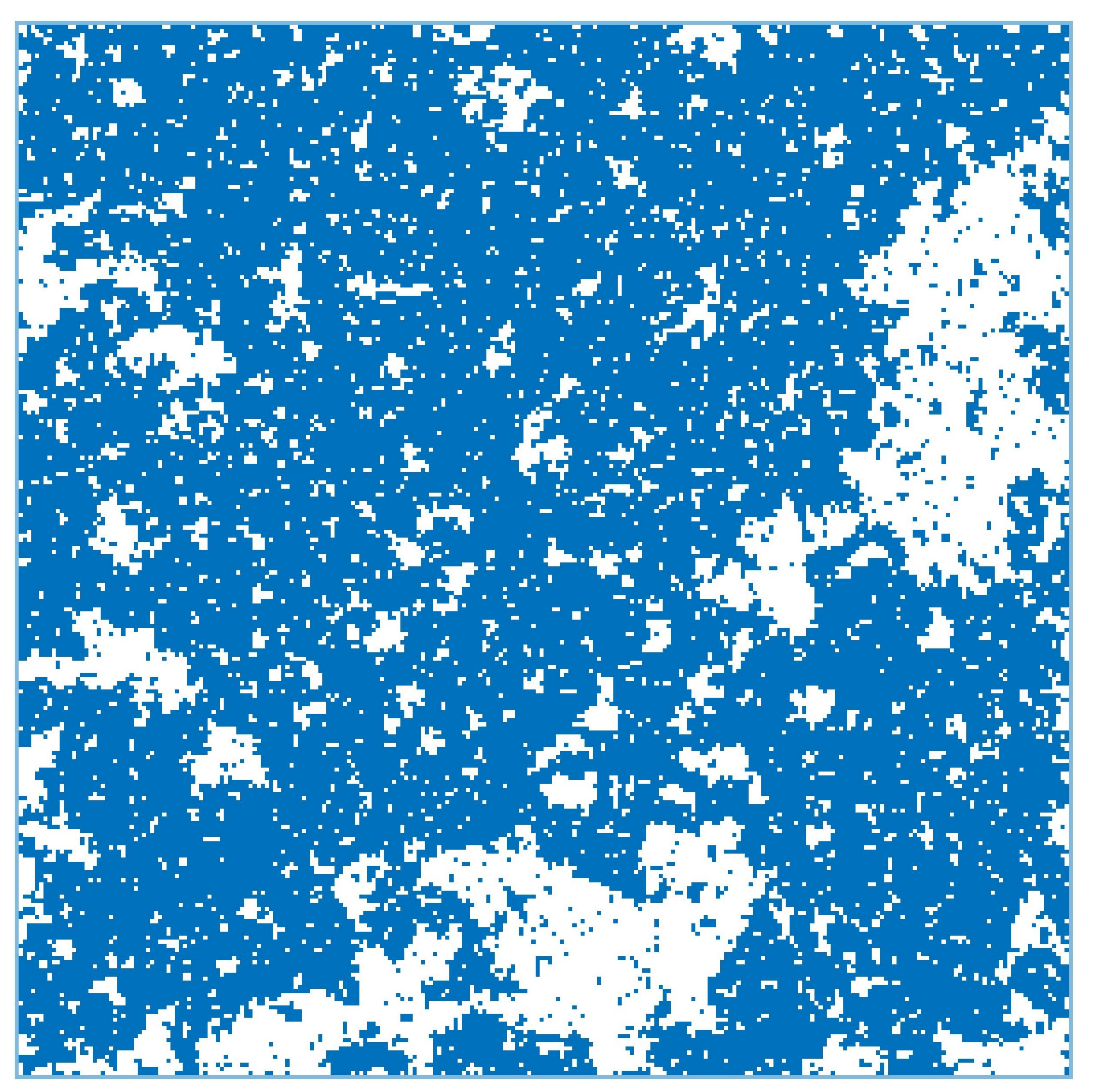}
		\caption{$T_{\text{2}} = 2.50$ , $T_1 \approx T_1^{c_1}$}
		\label{fig:b}
	\end{subfigure}
	\begin{subfigure}[b]{0.24\linewidth}
		\includegraphics[width=\linewidth]{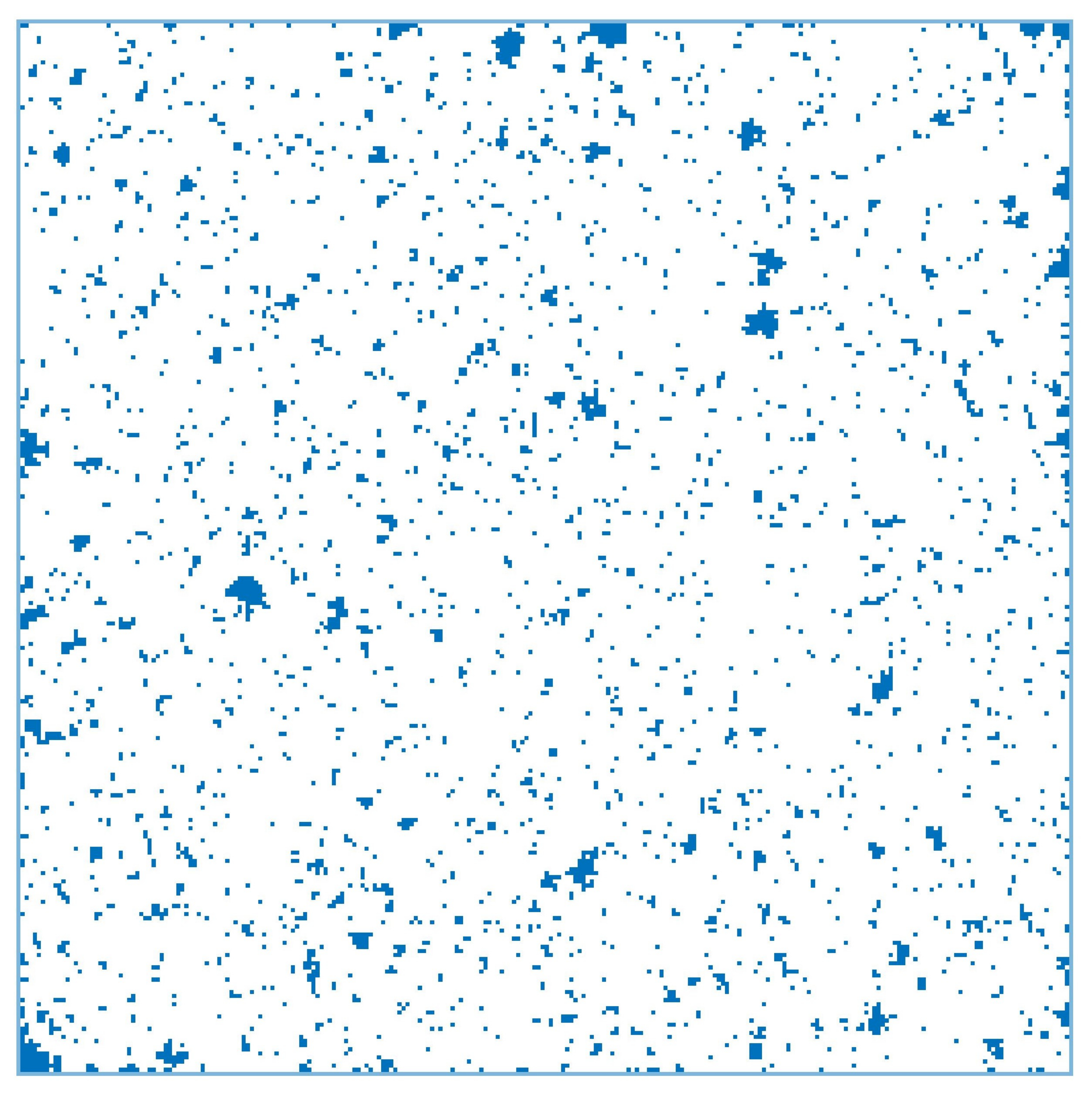}
		\caption{$T_{\text{2}} = 2.50$ , $T_1 = 1.50$}
		\label{fig:c}
	\end{subfigure}
	\begin{subfigure}[b]{0.24\linewidth}
		\includegraphics[width=\linewidth]{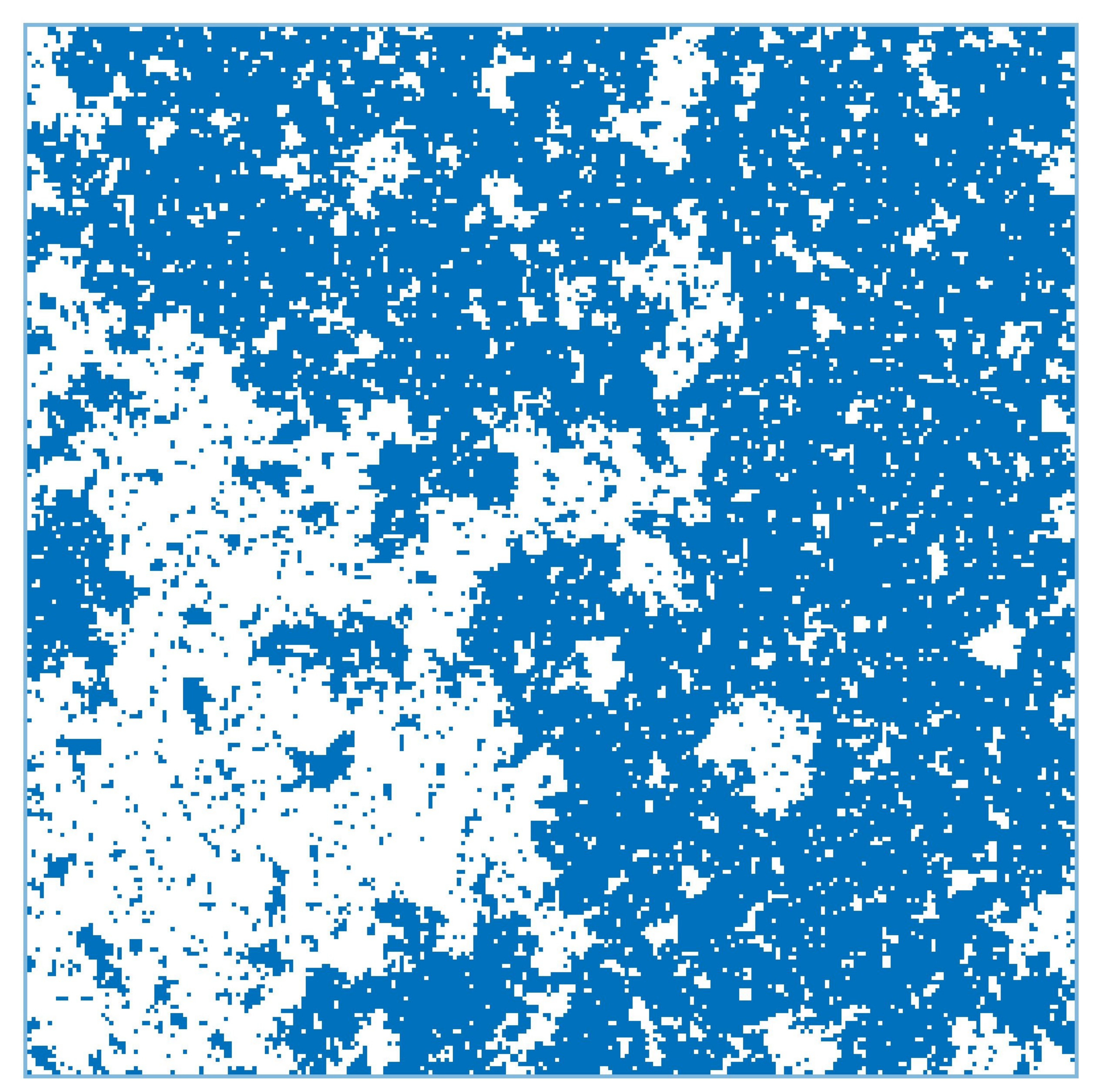}
		\caption{$T_{\text{2}} = 2.50$ , $T_1 \approx T_1^{c_2}$}
		\label{fig:d}
	\end{subfigure}
	\centering
	\caption{(Color online):The Ising samples by SW algorithm in $256 \times 256 $ lattice size at fixed  $T_{2} = 2.50$ for four variant temperatures $(T_1)$. For all fixed temperatures between regular Ising critical temperature ( $T_c \approx 2.26918$) and $T_D^{\models}$ there are two critical point. White sites show spin up and blue shows spin down}
	\label{tabcd}	
\end{figure*}

The other method that we use is the SW algorithm~\cite{swendsen1987nonuniversal}, in which instead of a single spin flip, a cluster, namely the Fortuin-Kasteleyn (FK) cluster~\cite{janke2004geometrical}, is chosen to be flipped~\cite{swendsen1987nonuniversal}. To define FK clusters, let us define first the geometric spin cluster, which is the connected cluster formed by sites with the same spins. The FK cluster is obtained by link dilution between the neighboring sites in the geometric spin cluster. This dilution is performed in ordinary (single temperature) Ising system using the following probability for establishing the links between neighboring sites $P_{\text{dialation}}=1-P_{\text{link}}$, where
\begin{align}
P_{\text{link}} = 1-e^{-2\beta_0}.
\label{q3}
\end{align}
The overall process is just like the Metropolis method, i.e. one starts from a random configuration, and the Monte Carlo steps continues until reaching stationary state. The SW algorithm is more appropriate in the vicinity of the critical points, where the fluctuations rise, causing the problem of critical slowing down. Practically, for the Metropolis method it is more appropriate to start in the high temperature limit, and reduce the temperature slowly, whereas one can generate the samples at any given temperature using the SW algorithm. Now let us consider two-temperature case with 2LBF, where instead of Eq.~\ref{q3}, we propose the following probability of adding a spin to the cluster
\begin{align}
P_{\text{link}}^s = \frac{1-e^{-2\beta_1}}{1+e^{2(\beta_1-\beta_2)}}+\frac{1-e^{-2\beta_2}}{1+e^{-2(\beta_1-\beta_2)}},
\label{q4}
\end{align}
which is obtained by combining Eqs.~\ref{q10} and~\ref{a1}. Its generalization for $q$-state $n$-temperature Potts model is presented in Appendix~\ref{ap}. One can easily check that this probability, and the probability presented in Eq.~\ref{q10} are not trivially related to the one-temperature counterparts. For example one might try to make connection with one-temperature system by fixing $\beta_2$, and changing $\beta_1$ leading the system to undergo an order-disorder transition. The probability measures are however very different from a one-temperature system in this case, meaning that one cannot define an effective temperature in one-temperature system with a well-define Boltzmann factor equivalent to this 2LBF system, which makes the properties of the system different and non-trivial.\\
In this paper we applied both algorithms for comparison reasons. Some samples that were obtained using these algorithms are shown in Fig.~\ref{tabcd} (using the Metropolis algorithm), and~\ref{t56} (using the SW algorithm). 
We observed that this system undergoes an order-disorder transition which defines an extended critical line. Let us define $T_c\equiv \left( T_1^c,T_2^c\right) $ as the critical points that this transition takes place. In obtaining the phase diagram we fix one temperature (say $T_2$) and change the other one ($T_1$). We will see that there are three possibilities depending on the chosen value of $T_2$: one may see zero, one and two transition points (see the following section). Let us define the transition points by $T_{1,2}^{c_i}$ as the $i$th critical temperature for fixed $T_{2,1}$, where depending on the imposed conditions there are zero, one ($i\in \left\lbrace 1\right\rbrace $) or two ($i\in \left\lbrace 1,2\right\rbrace$) transition points. The measures/observables for detecting criticality and extracting exponents are the heat capacity
\begin{align}
C_v = N( \langle E^2 \rangle - {\langle E \rangle}^2),
\end{align}
where $\langle \rangle$ represents ensemble average. The other quantity is the Binder’s cumulant also known as the fourth-order cumulant
\begin{equation}
U_4 = 1-\frac{\langle m^4 \rangle }{3 {\langle m^2 \rangle}^2},
\label{q5}
\end{equation}
where ${\langle m^4 \rangle }$ is fourth moment of magnetization ($m\equiv\frac{1}{N}\sum_{i=1}^{N}s_i$ for each sample, and $N=L^2$ is the number of sites in the sample) and ${\langle m^2 \rangle}$ is second moment of the magnetization. It has been defined as the kurtosis of the order parameter. The phase transition point is usually identified comparing the behavior of $U_4$ as a function of the temperature for different values of the system size $L$.\\

Also the magnetic susceptibility which is defined by
\begin{equation}
\chi =N (\langle m^2 \rangle - {\langle m \rangle}^2),
\label{q6}
\end{equation}
is calculated in this work which is expected to diverge in a power-law fashion at the transition point for the ordinary Ising model.

\section{Numerical results}\label{Sec:num}
\begin{figure}
	\centerline{\includegraphics[scale=0.6]{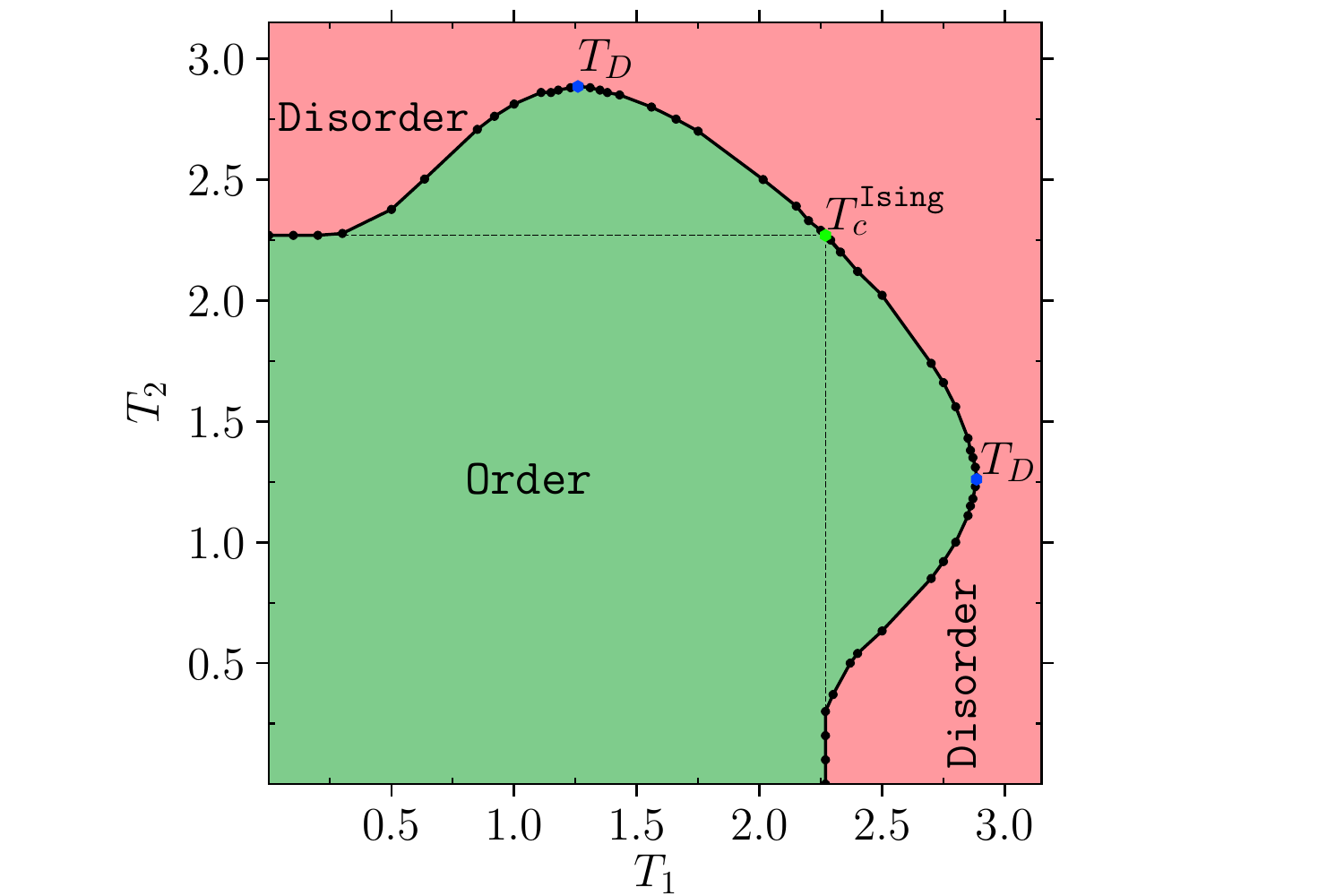}}
	\caption{ The phase diagram for $T_1$ and $T_2$. The read area indicate disorder phase, and the green area indicate order phase}
	\label{phase}
\end{figure}

In our simulations we fix $T_2$, and run the program (for both Metropolis and SW algorithms) starting from high temperatures. The ensemble averages were done upon $10^{4}$ samples. Before going to details, let us summarize the main results which facilitate reading the rest of the paper. The phase diagram is shown in Fig.\ref{phase}, were the bold circles are the transition points obtained by simulations, and the black bold line is the interpolation between points to help eye. At $T_2=0$, the system undergoes a continuous transition at $T_1^c\approx 2.269$ as expected for the Ising model on the square lattice. Also one can distinguish a $T_1\leftrightarrow T_2$ symmetry in this phase diagram as expected. We see from this figure that in the $T_2$ direction there is a highest point $T_D\equiv (T_D^{\vdash},T_D^{\models})$ (which we estimated them to be $T_D^{\vdash} = 1.270 \pm 0.003$ and $T_D^{\models} = 2.885 \pm 0.003$) which separates the properties of the model. More precisely, when we fix $T_2$, for $T_2<T_c^{\text{Ising}}$ ($T_c^{\text{Ising}}$ being the critical temperature for ordinary two-dimensional Ising model) we have one second-order transition point for $T_1$, and for $T_c^{\text{Ising}}\leq T_2< T_D^{\models}$ we have two second-order transition points for $T_1$, and for $T_2=T_D^{\models}$ these two transition points merge so that we have one tricritical point for $T_1$, and for $T_2>T_D^{\models}$ there is no transition. In the remaining we characterize this transition. 
The same features are seen in terms of $T_1$, i.e. if we increase $T_2$ slowly from zero, the critical point starts to increase and gets away from $T_c^{\text{Ising}}$ up to the point $T_D^\vdash$, after which the critical point stars to decrease until $T_{2} = T_c^{\text{Ising}}$.\\

\begin{figure*}
	\begin{subfigure}{0.49\textwidth}\includegraphics[width=\textwidth]{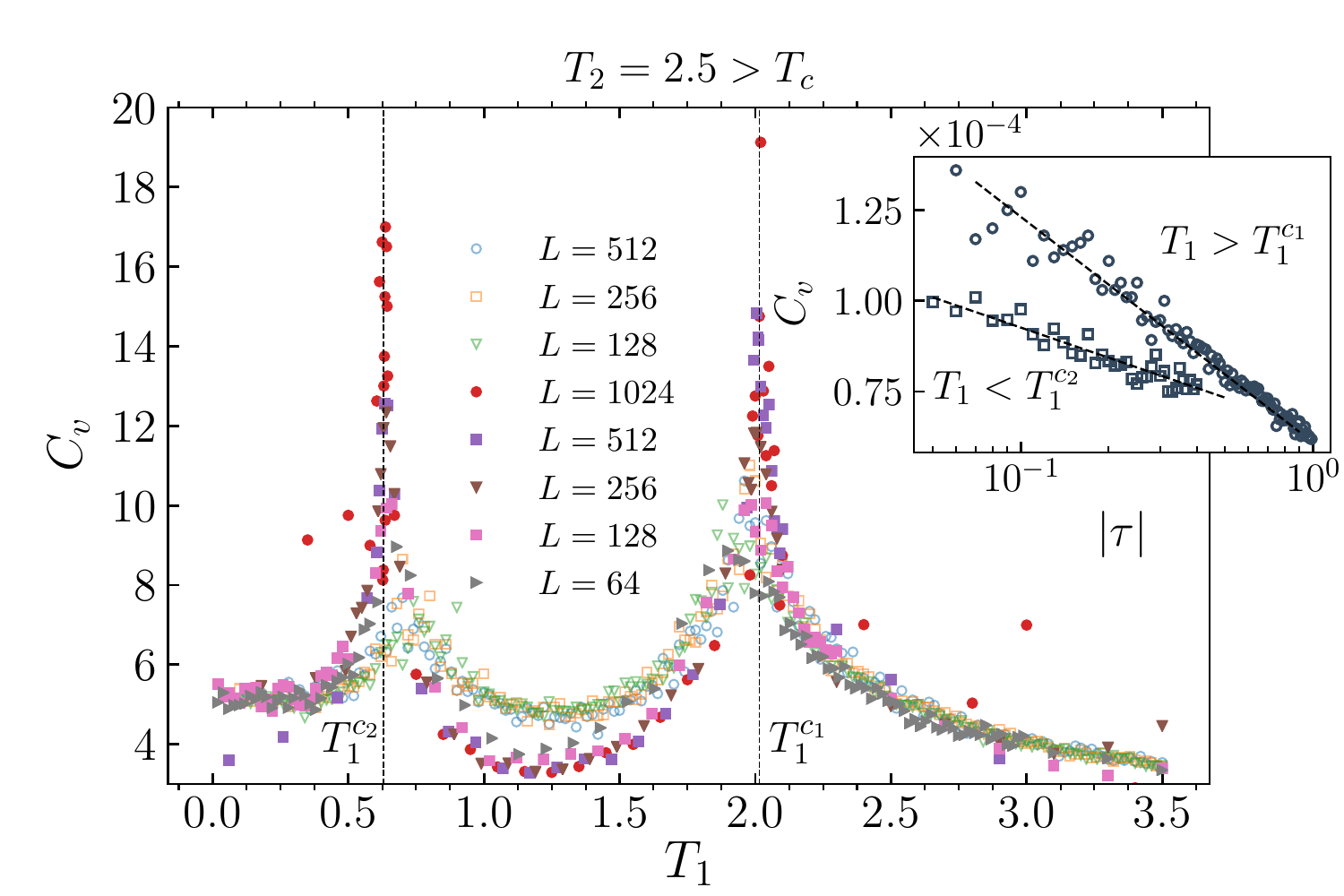}
		\caption{}
		\label{cv1}
	\end{subfigure}
	\centering
	\begin{subfigure}{0.49\textwidth}\includegraphics[width=\textwidth]{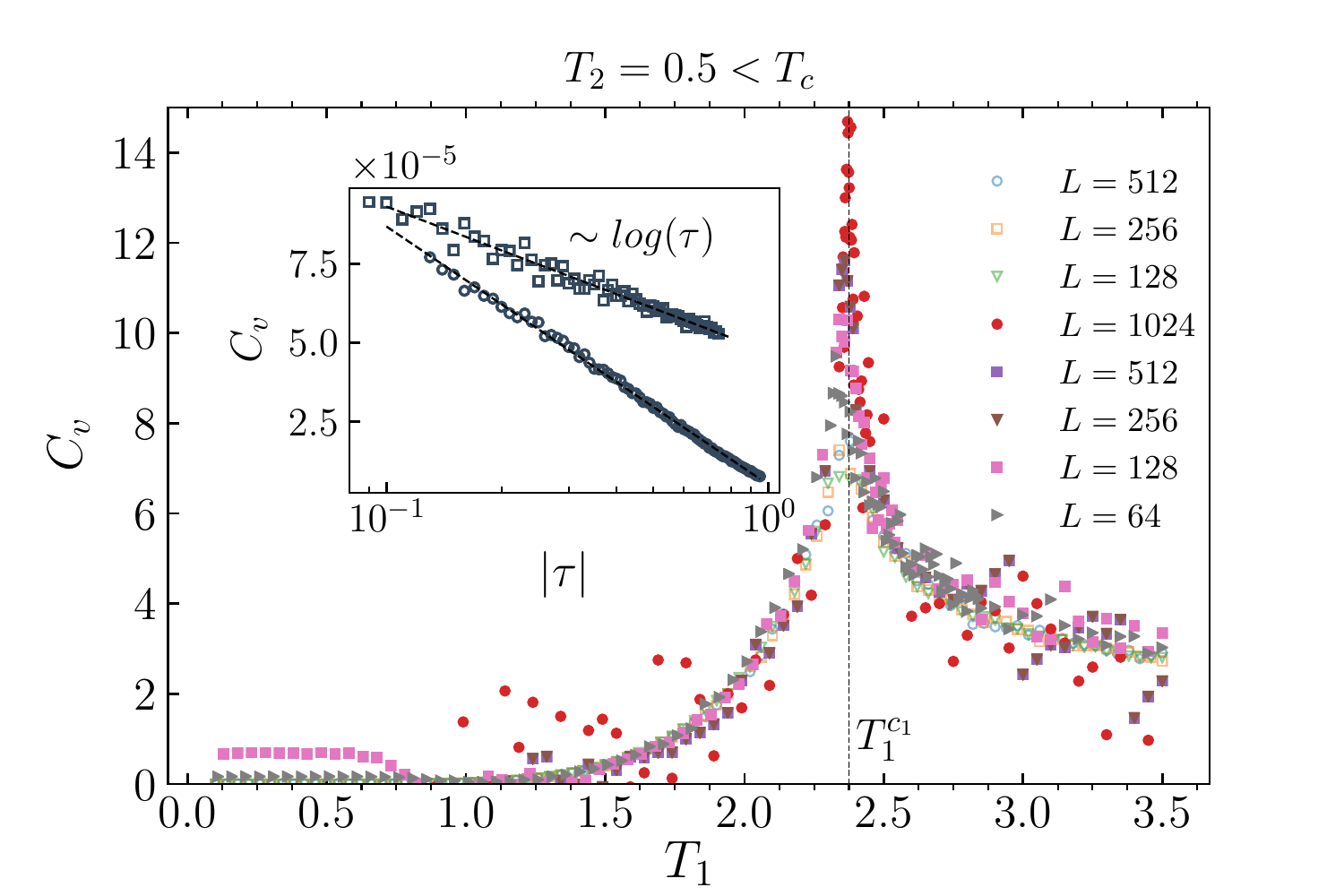}
		\caption{}
		\label{cv2}
	\end{subfigure}
	\caption{(Color online): (a) Heat capacity ($C_v$) for various system size in fixed $T_2 = 2.5>T_c $. Inset: The critical exponent for $C_v$ is logarithmic scale for $T_1>T_1^{c_2}$ and $T_1<T_1^{c_1}$ where  $\alpha =0$.  (b) Heat capacity ($C_v$) for various system size in fixed $T_2 = 0.50<T_c $. Inset: The critical exponent for $C_v$ is logarithmic scale for $T_1>T_1^{c_1}$ and $T_1<T_1^{c_1}$ where  $\alpha =0$. [Filled marker represent SW algorithm and hollow marker Metropolis] }
	\label{Cv}
\end{figure*}

The results for the heat capacity are shown in figs.~\ref{cv1} and~\ref{cv2} at fixed $T_2 =2.50>T_c^{\text{Ising}}$ and $T_2 =0.50<T_c^{\text{Ising}}$ in terms of $T_1$ for various system sizes. As claimed above we see that two second order phase transitions occur at $T_1^{c_1}=2.015 \pm 0.005$ and $T_1^{c_2}=0.630 \pm 0.005$ when $T_2 =2.50>T_c^{\text{Ising}}$ (fig.~\ref{cv1}), and a second-order phase transition at fixed temperature $T_2 =0.50<T_c$ at $T_1^{c_2}=2.375 \pm 0.005$ (fig.~\ref{cv2}). According to the inset of fig.~\ref{cv1}, this function behaves logarithmically in terms of $\tau$, ($\tau = T_1 - T_1^{c_1}$, for $T_1>T_1^{c_1}$ and $\tau = T_1^{c_2}-T_1$ for $T_1<T_1^{c_2}$), so that the $\alpha$ exponent in two case is zero the same as the regular Ising model. Fig.~\ref{cv2} shows similarly that $\alpha$ exponent for $T_2 = 0.50$ is zero, ether for $T_1 > T_1^{c_1}$ or $T_1 < T_1^{c_1}$. For comparison two different simulation methods we have shown the results for both SW and Metropolis Monte Carlo methods.\\

\begin{figure*}
	\begin{subfigure}{0.49\textwidth}\includegraphics[width=\textwidth]{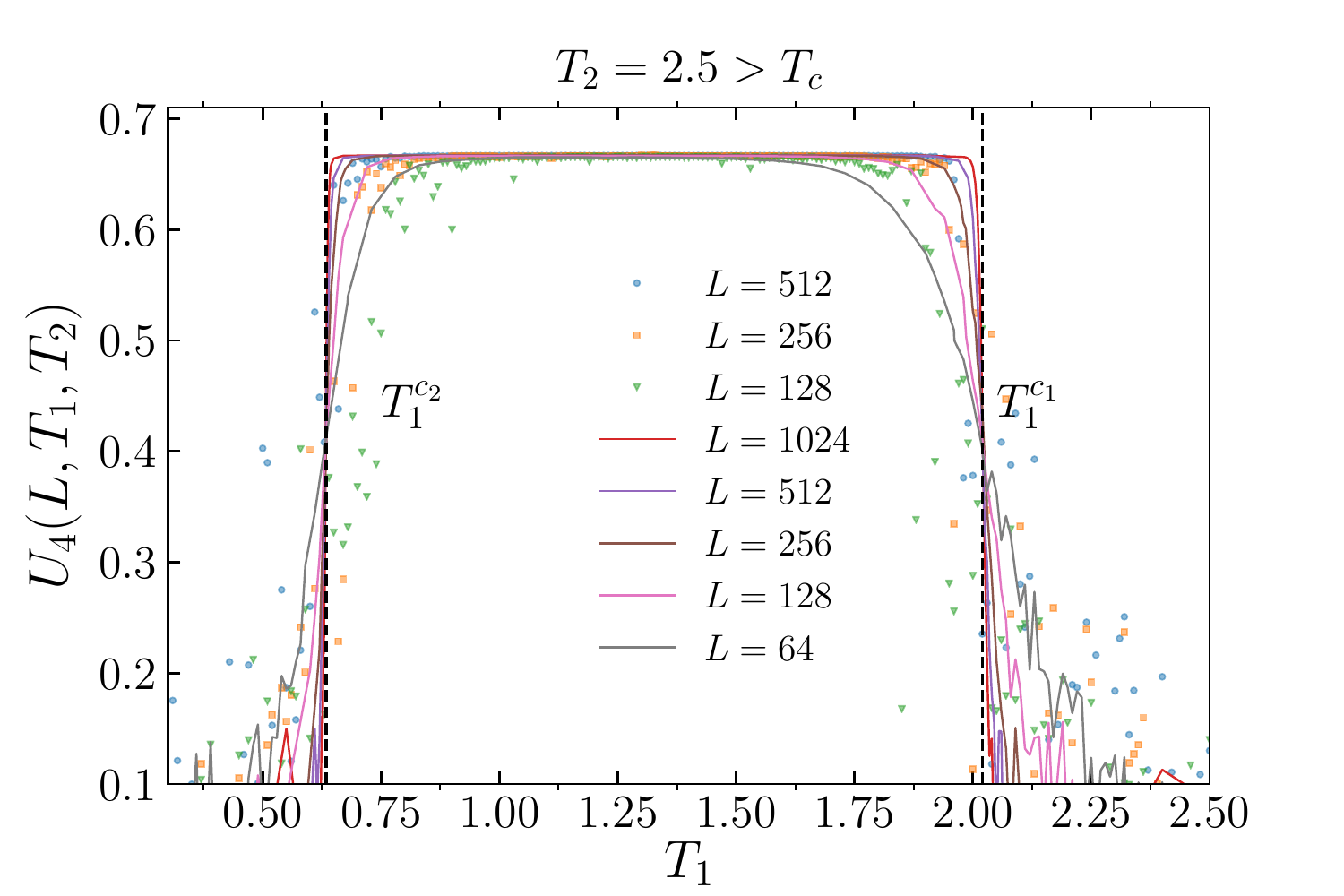}
		\caption{}
		\label{fig:u}
	\end{subfigure}
	\begin{subfigure}{0.49\textwidth}\includegraphics[width=\textwidth]{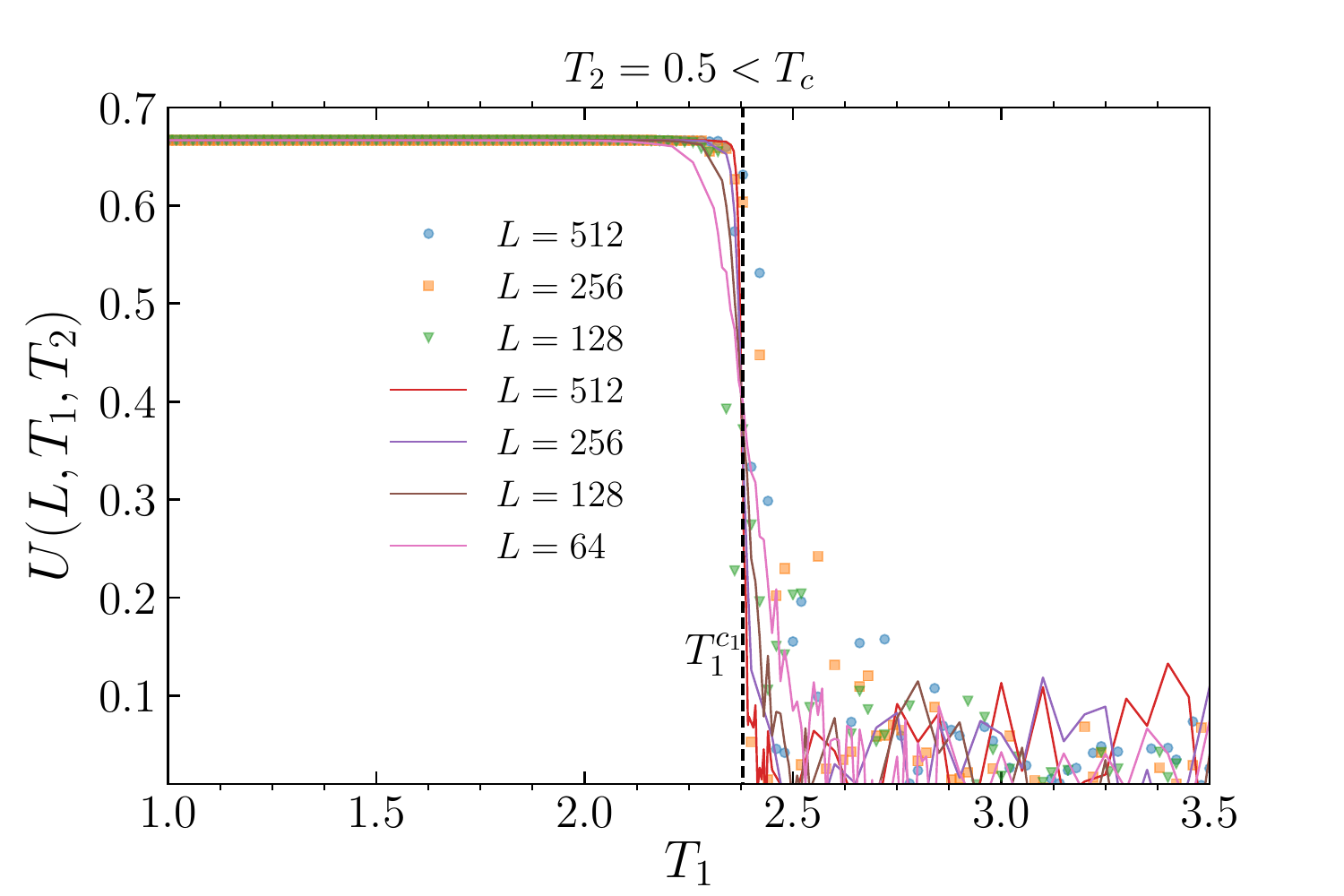}
		\caption{}
		\label{fig:u1}
	\end{subfigure}
	\caption{(Color online): (a) Binder’s cumulant($U_4$) in terms of $T_1$  for various system size in fixed $T_2 = 2.50>T_c $. (b) Binder’s cumulant for various system size in fixed $T_2 = 0.50<T_c $. [Solid line represent SW algorithm and hollow marker is Metropolis method.] }
	\label{pashe}
\end{figure*}

The Binder cumulant analysis is shown in Fig.(\ref{fig:u})  and Fig.(\ref{fig:u1}) in terms of $T_1$ again in fixed $T_2 = 2.50$ and $T_2 = 0.50$ respectively. These figures confirm the results of heat capacity, i.e. two critical points are seen for fixed temperature $T_2 = 2.50$ at  $T_1^{c_1}=2.020 \pm 0.005$ and $T_1^{c_2}=0.634 \pm 0.005$, and one critical point for  $T_2 = 0.50 $ at $T_1^{c_2}=2.380 \pm 0.005$. In these figures solid line shows SW algorithm and symbols represent the Metropolis method.\\

\begin{figure*}
	\begin{subfigure}{0.49\textwidth}\includegraphics[width=\textwidth]{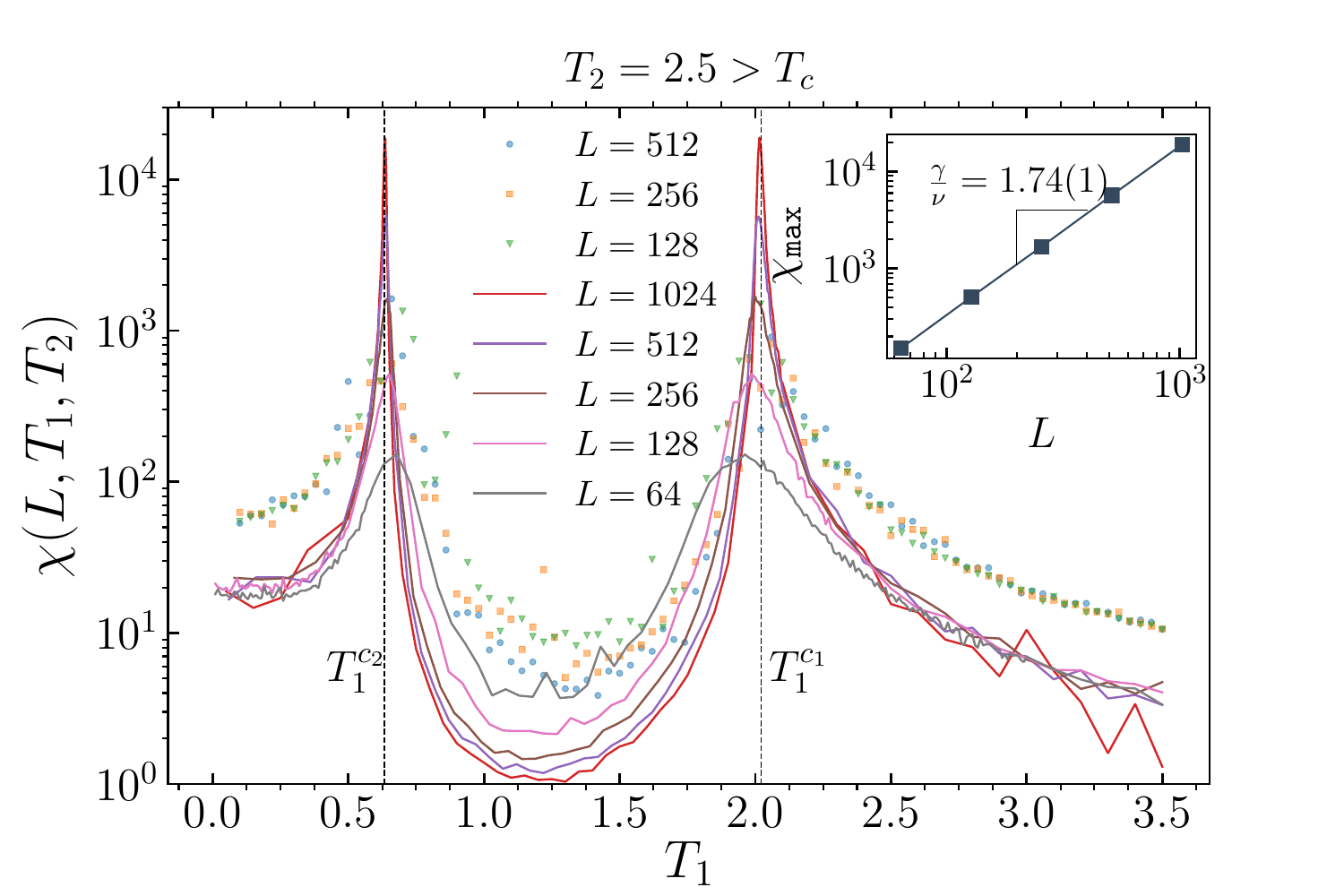}
		\caption{}
		\label{fig:xi}
	\end{subfigure}
	\centering
	\begin{subfigure}{0.49\textwidth}\includegraphics[width=\textwidth]{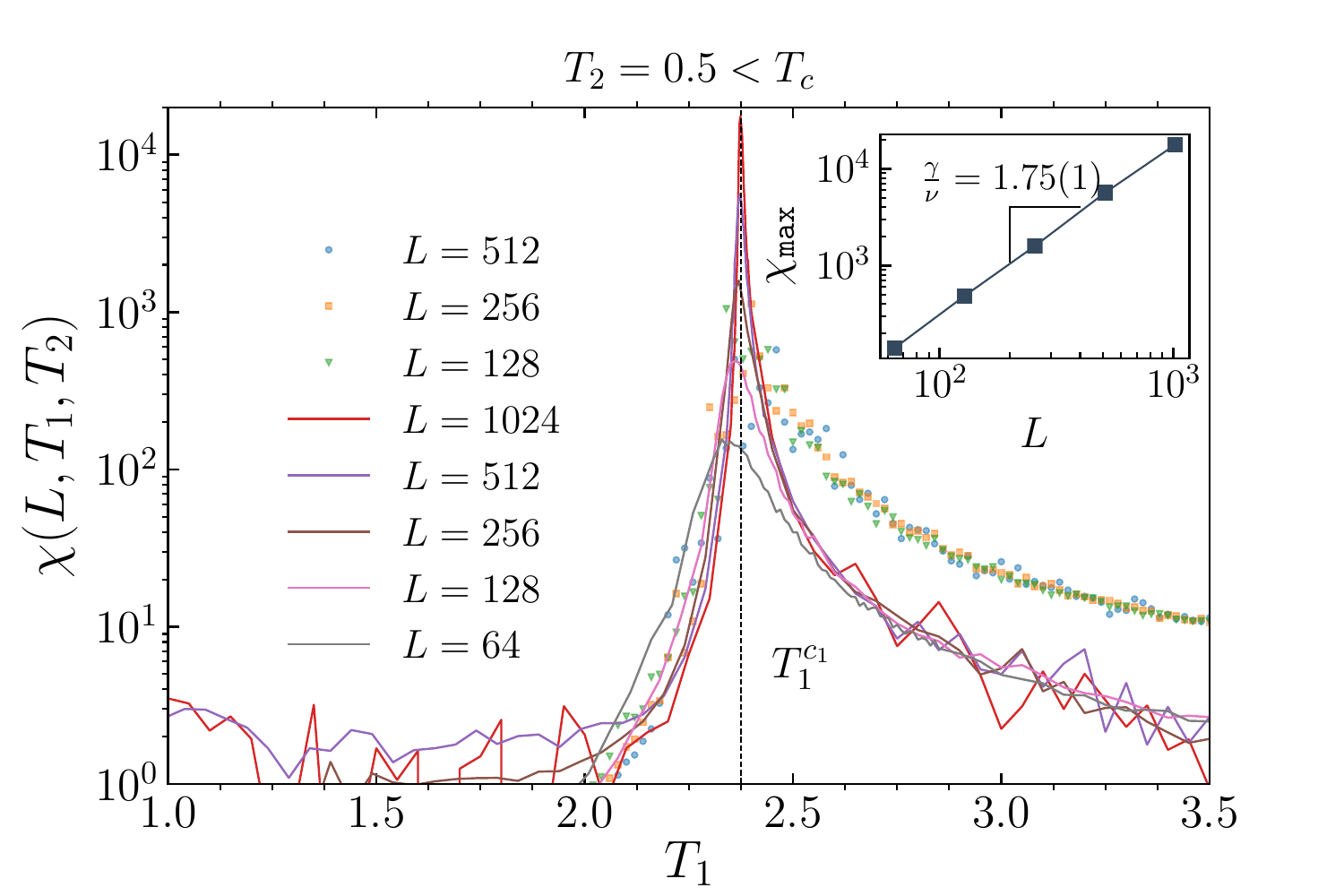}
		\caption{}
		\label{fig:xi1}
	\end{subfigure}
	\centering
	\caption{(Color online): (a) Magnetic susceptibility ($\chi$) for various system size in fixed $T_2 = 2.50>T_c $. Inset:$\chi_{\text{max}}$ in term of lattice size in which show $\gamma /\nu $ exponent (b) Magnetic susceptibility in fixed $T_2 = 0.50<T_c $. Inset: $\chi_{\text{max}}$ in term of lattice size in which show $\gamma /\nu $ exponent. [Solid line represent SW algorithm and hollow marker is Metropolis method.] }
	\label{xi}
\end{figure*}

The magnetic susceptibility is shown in figs.~\ref{fig:xi} and ~\ref{fig:xi1} for various system size in fixed $T_2 = 2.50$ and $T_2 = 0.50$ respectively where the similar results are obtained, i.e. $T_1^{c_1}=2.015 \pm 0.005$ and $T_1^{c_2}=0.633 \pm 0.005$ for he first case, and $T_1^{c_2}=2.375 \pm 0.005$ for the latter case. Scaling hypothesis predicts that the maximum value of $\chi$ at the transition point behaviors like $\chi_{max} \sim L^{\gamma / \nu}$. The insets of Fig.(\ref{xi})  shows that $\gamma / \nu = 1.75 \pm 0.01$. It is equal to Ising critical exponent ($\gamma / \nu = 7/4$). We observed that this is the case (for all exponents that we found in this paper) for all transition points for temperatures bellow $T_D^{\models}$.\\

\begin{figure*}
	\begin{subfigure}{0.49\textwidth}\includegraphics[width=\textwidth]{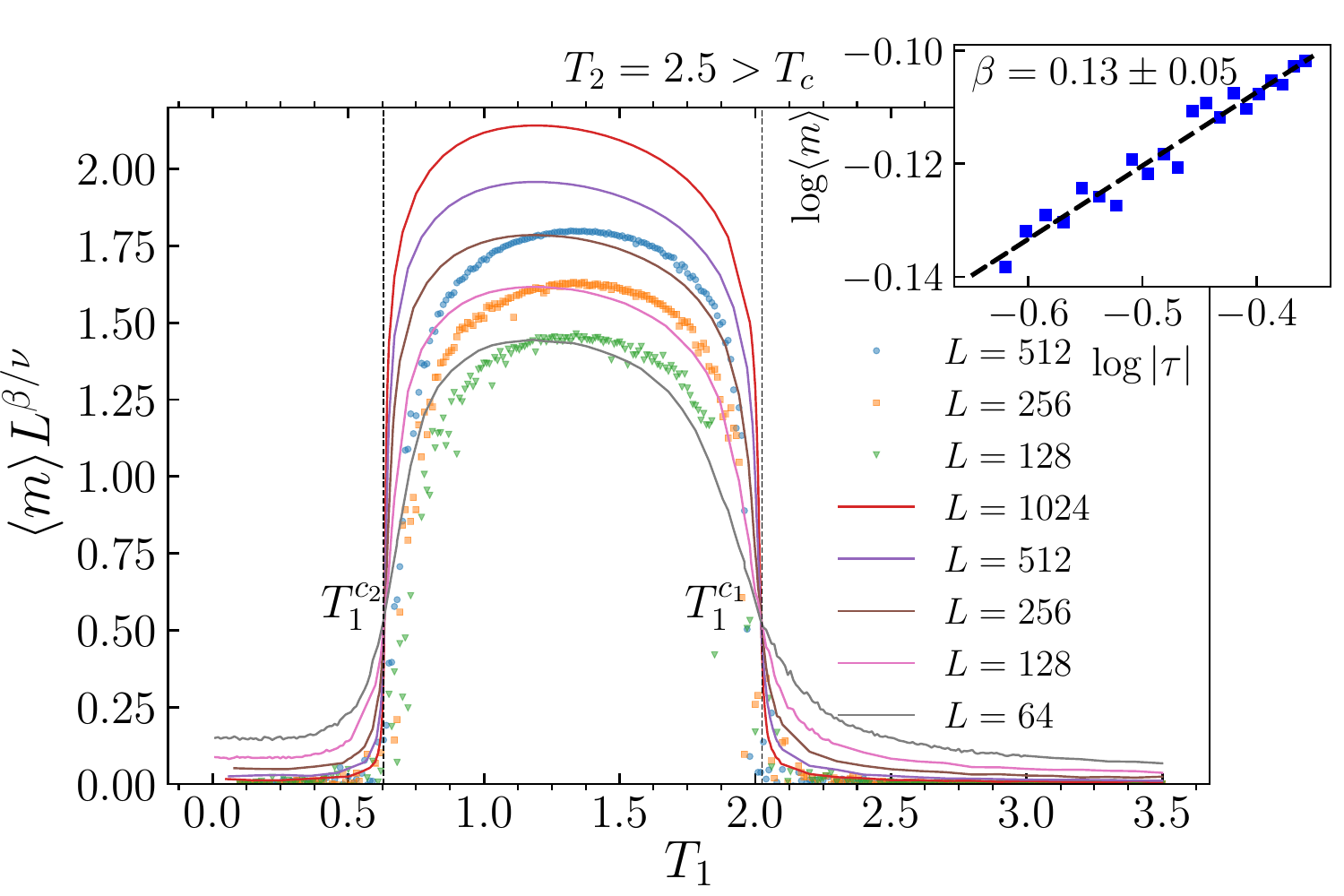}
		\caption{}
		\label{fig:m}
	\end{subfigure}
	\centering
	\begin{subfigure}{0.49\textwidth}\includegraphics[width=\textwidth]{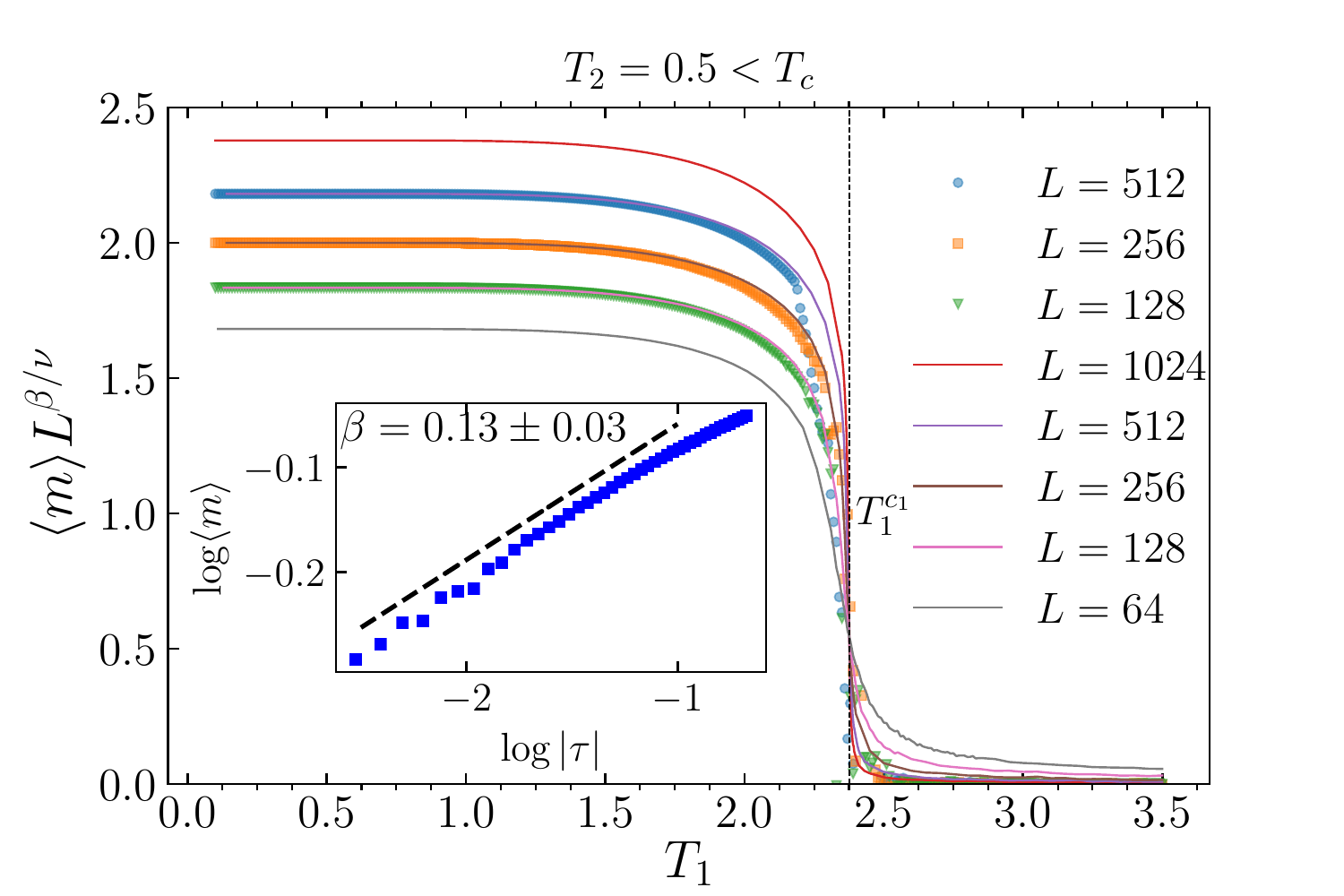}
		\caption{}
		\label{fig:m1}
	\end{subfigure}
	\centering
	\caption{(Color online): (a) $\langle m \rangle L^{\beta / \nu}$  in terms of $T_1$ for various system sizes in fixed $T_2 = 2.50>T_c $. inset : $\langle m \rangle \sim |\tau|^{\beta}$ in which $\beta = 0.13 \pm 0.05$.  (b) $\langle m \rangle L^{\beta / \nu}$ in terms of $T_1$ for various system sizes in fixed $T_2 = 0.50<T_c $. inset:$\langle m \rangle \sim |\tau|^{\beta}$ in which $\beta = 0.13 \pm 0.3$. [Solid line represent SW algorithm and hollow marker is Metropolis method.] }
	\label{mm}
\end{figure*}

The scaling behavior of $\left\langle m\right\rangle $ gives us some other important exponents. By tracking the behavior of this function in terms of $T_1$ and $L$ (for fixed $T_2$), one can extract the critical temperature $T_1^{c_{1,2}}$, as done in the  fig.~\ref{mm}, in which for a best choose of $\beta/ \nu$ all curves cross each other in a critical point, defined via the following scaling relation
\begin{equation}
m(\epsilon)=L^{-\beta/\nu}G_m(\epsilon L^{1/\nu}),
\label{Eq:scaling1}
\end{equation}
where $\epsilon\equiv \frac{T_1-T_1^{c_{1,2}}}{T_1^{c_{1,2}}}$, $G_m(x)$ is a scaling function with $G_m(x)|_{x\rightarrow \infty}\propto x^{\beta}$ and is analytic and finite as $x\rightarrow 0$. From this analysis we observe recover the above results for the critical temperatures. By plotting $\langle m \rangle L^{\beta/ \nu}$ in terms of $T_1$, we find that $\beta/ \nu = 0.13 \pm 0.05$ and $\nu = 1.00 \pm 0.05 $, just the same as the regular Ising model. This result is correct for all temperature on the critical line.

\begin{figure*}
	\begin{subfigure}[b]{0.49\linewidth}	
		\centering
		\includegraphics[width=\linewidth]{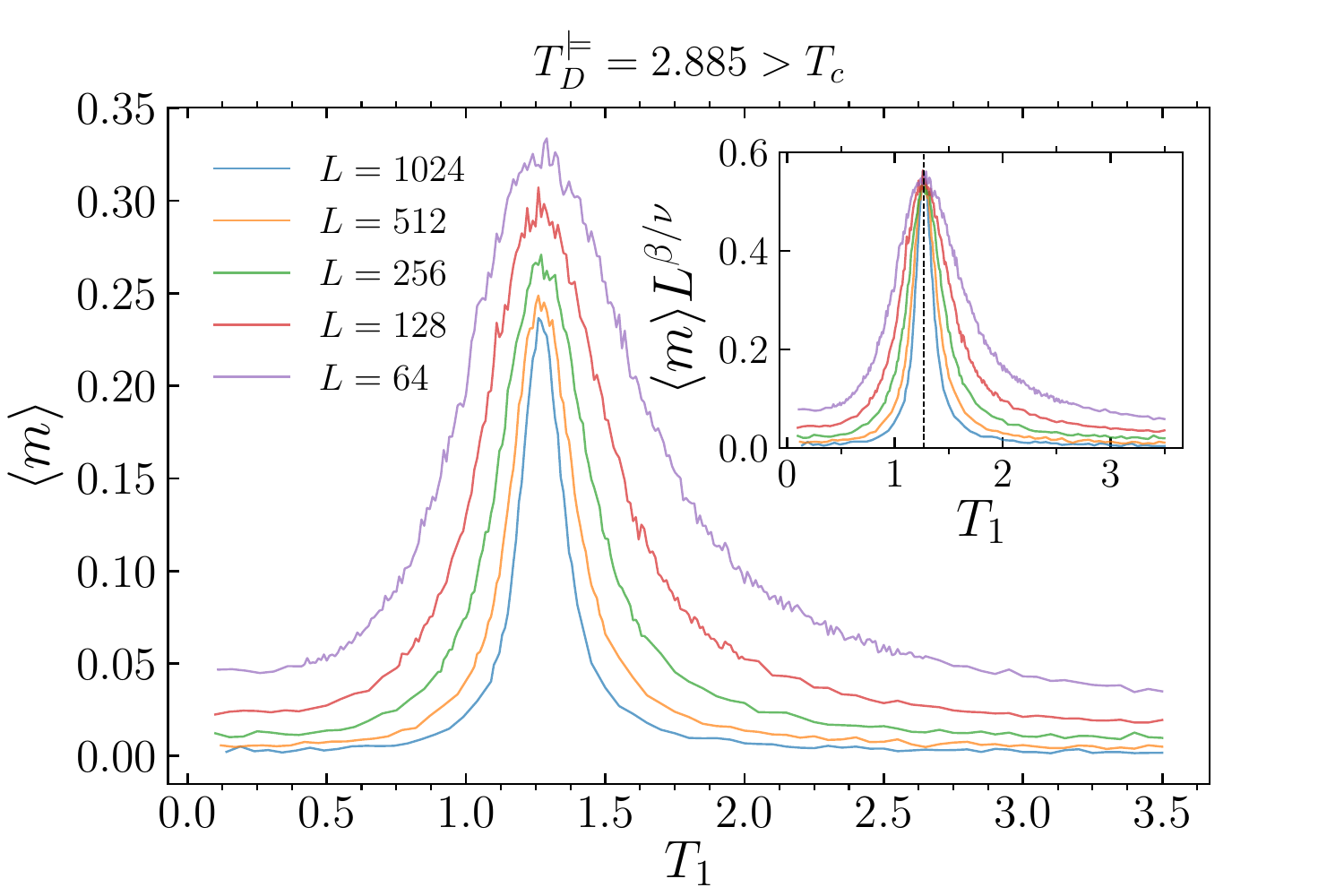}
		\caption{}
		\label{fig:md}
	\end{subfigure}
	\begin{subfigure}[b]{0.49\linewidth}
		\centering
		\includegraphics[width=\linewidth]{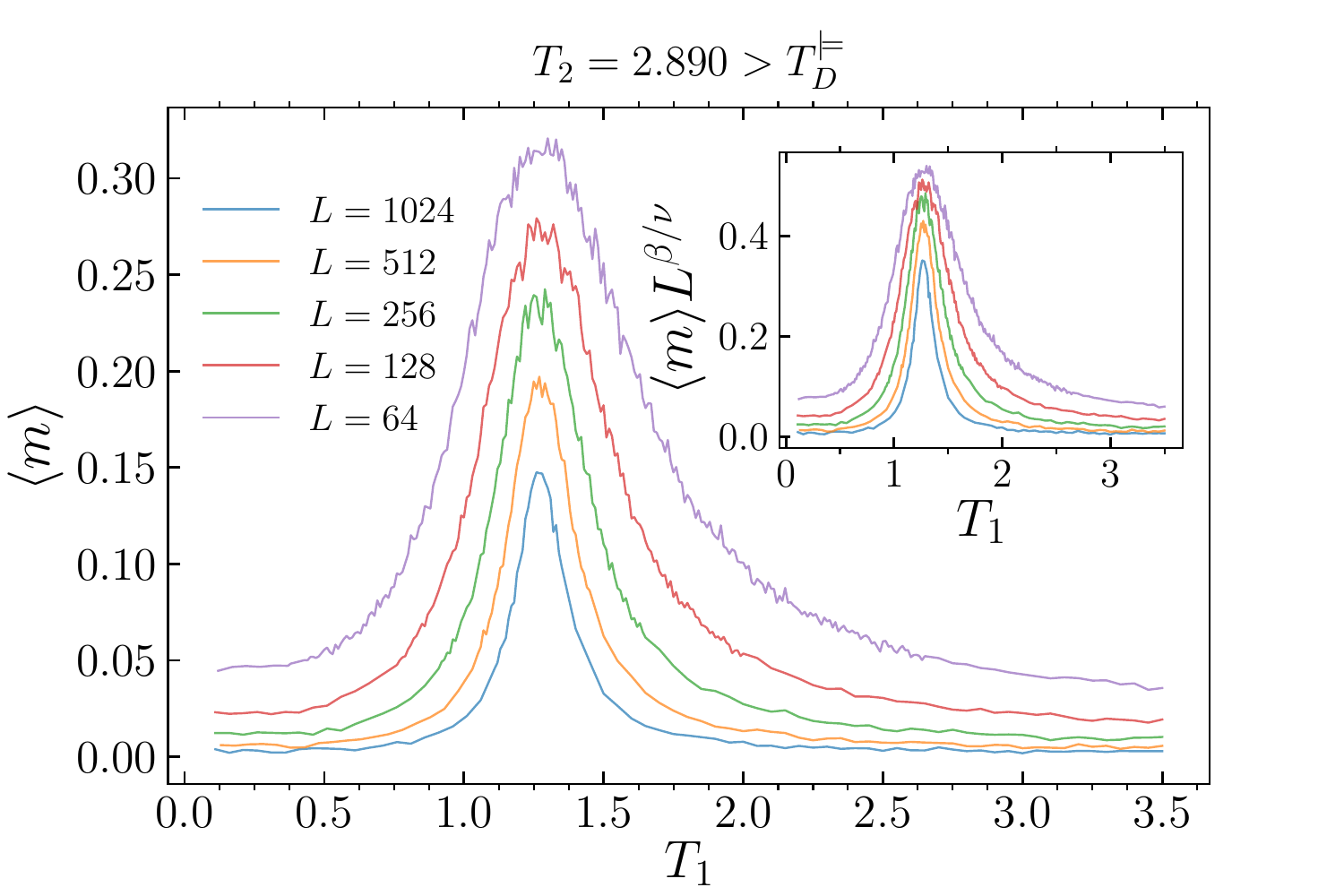}
		\caption{}
		\label{fig:dmd}		
	\end{subfigure}
	\centering
	\caption{(Color online):(a) $\langle m \rangle$ in term of $T_1$ for various system size in fixed $T_D^{\models} = 2.885>T_c $. Inset: $\langle m \rangle L^{\beta / \nu}$  in term of $T_1$ for various system size. (b)$\langle m \rangle$ in term of $T_1$ for various system size in fixed $T_2 = 2.890>T_D^{\models} $ inset : $\langle m \rangle \sim |\tau|^{\beta}$ in which curve does not cross each other. }
	\label{t12}	
\end{figure*}

\begin{figure*}
	\begin{subfigure}[b]{0.49\linewidth}	
		\centering
		\includegraphics[width=\linewidth]{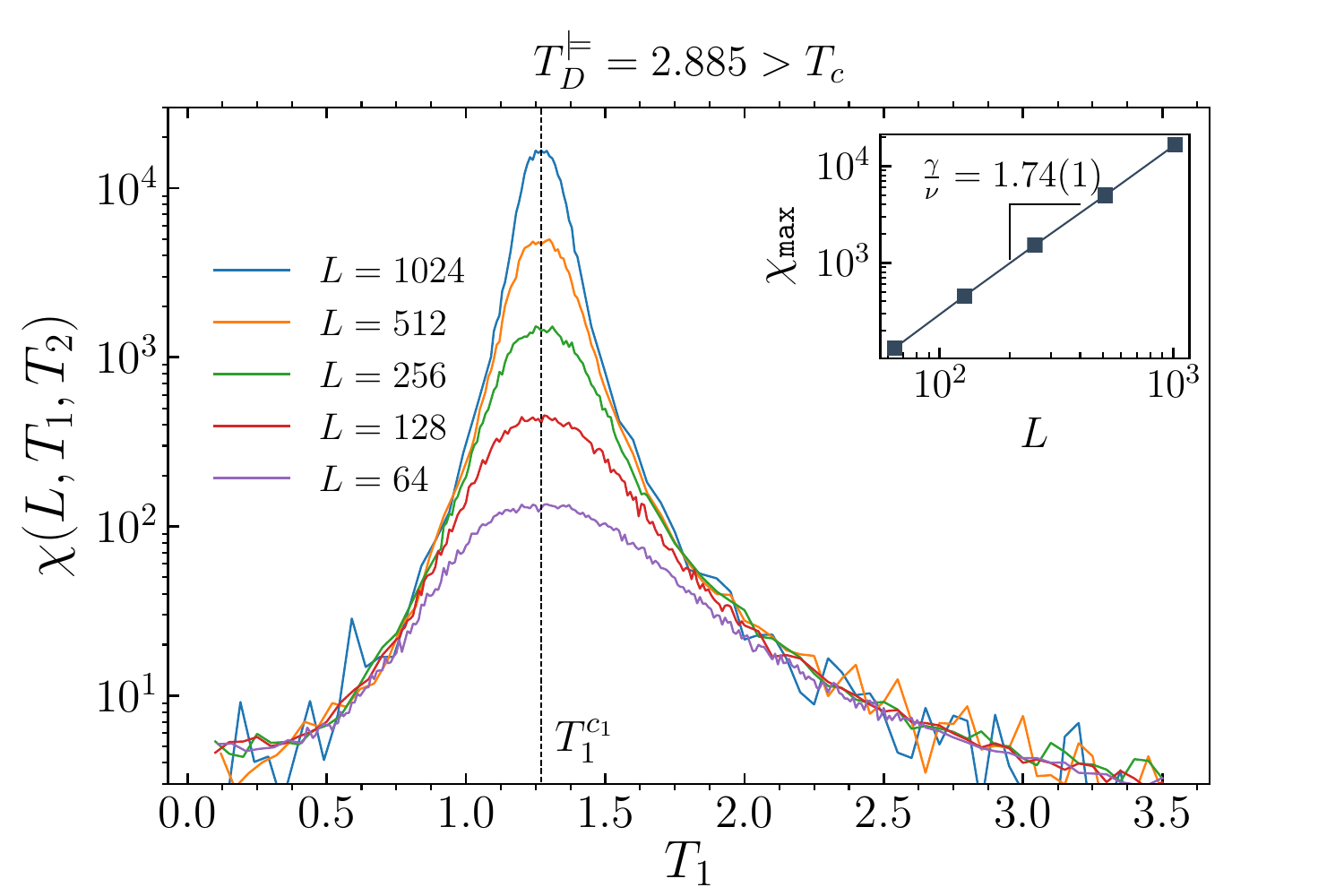}
		\caption{}
		\label{fig:Xid}
	\end{subfigure}
	\begin{subfigure}[b]{0.49\linewidth}
		\centering
		\includegraphics[width=\linewidth]{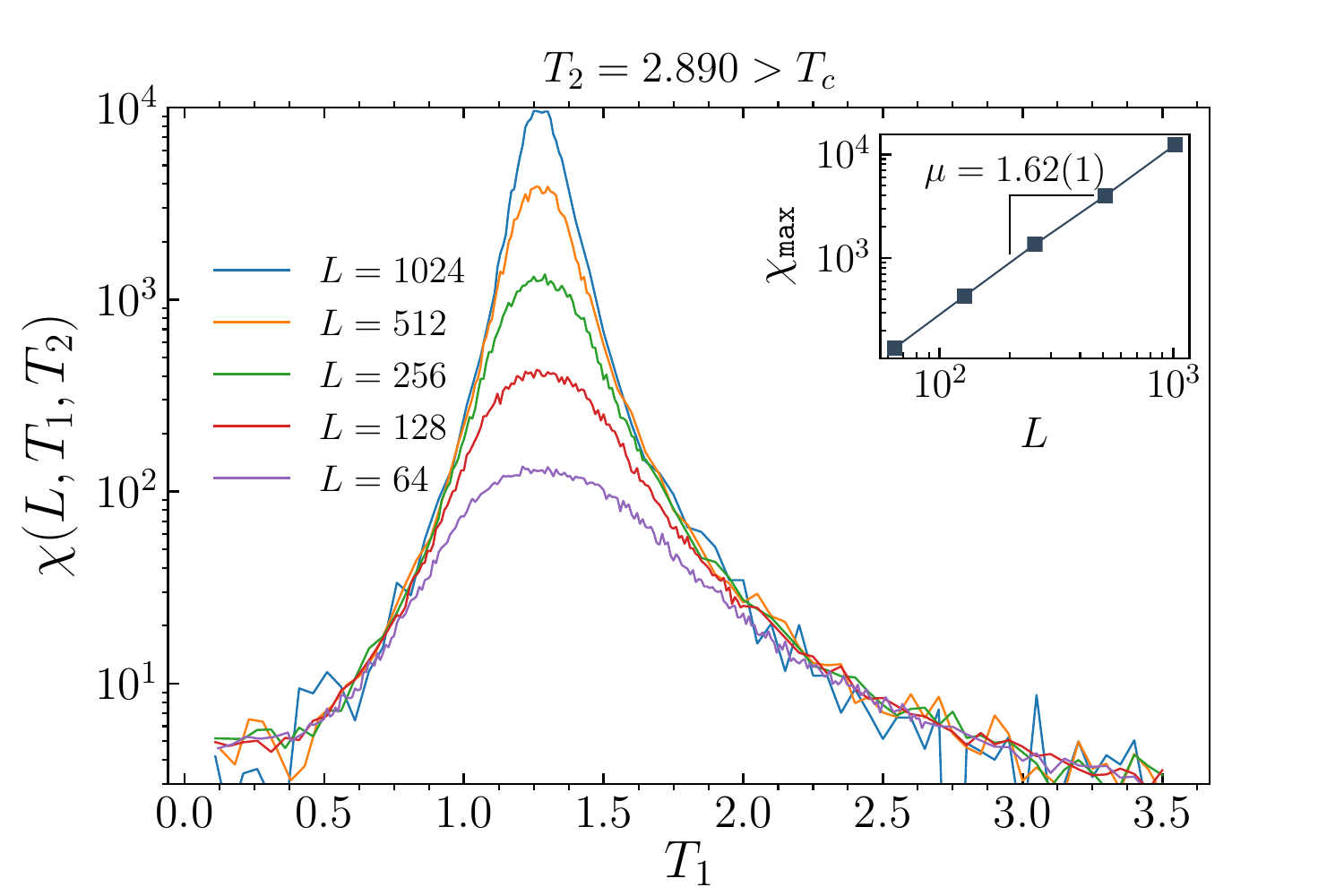}
		\caption{}
		\label{fig:DXid}		
	\end{subfigure}
	\centering
	\caption{(Color online): (a) Magnetic susceptibility ($\chi$) for various system size in fixed $T_D^{\models} = 2.885>T_c $. Inset:$\chi_{\text{max}}$ in term of system size in which $\gamma /\nu = 1.74 \pm 0.01 $. (b) Magnetic susceptibility ($\chi$) for various system size in fixed $T_2 = 2.890>T_D^{\models} $. Inset:$\chi_{\text{max}}$ in term of lattice size in $\mu = 1.62 \pm 0.01 $ . }
	\label{t12}	
\end{figure*}

\begin{figure*}
	\begin{subfigure}[b]{0.50\linewidth}	
		\centering
		\includegraphics[width=\linewidth]{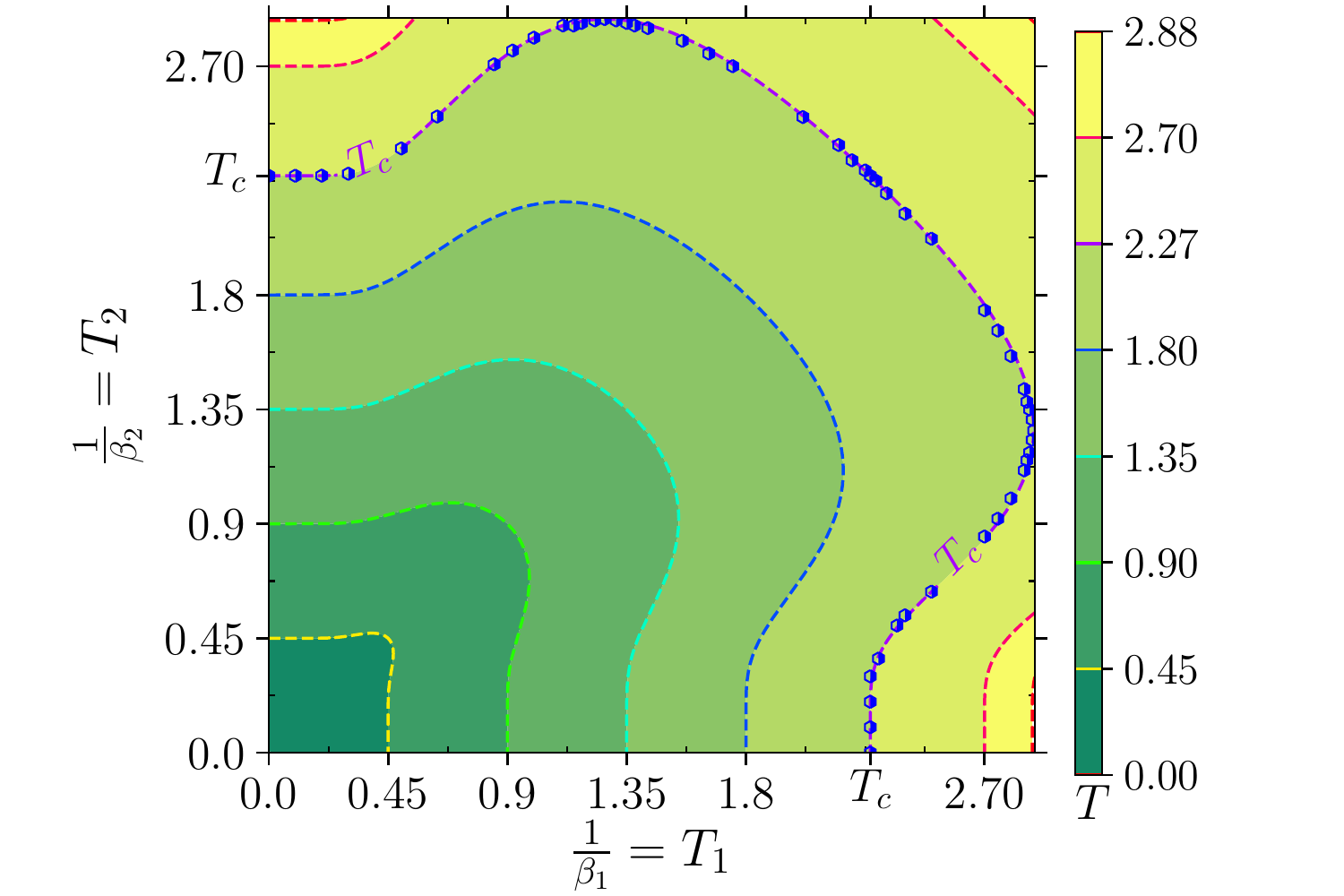}
		\caption{}
		\label{fig:plnk}
	\end{subfigure}
	\begin{subfigure}[b]{0.49\linewidth}	
		\centering
		\includegraphics[width=\linewidth]{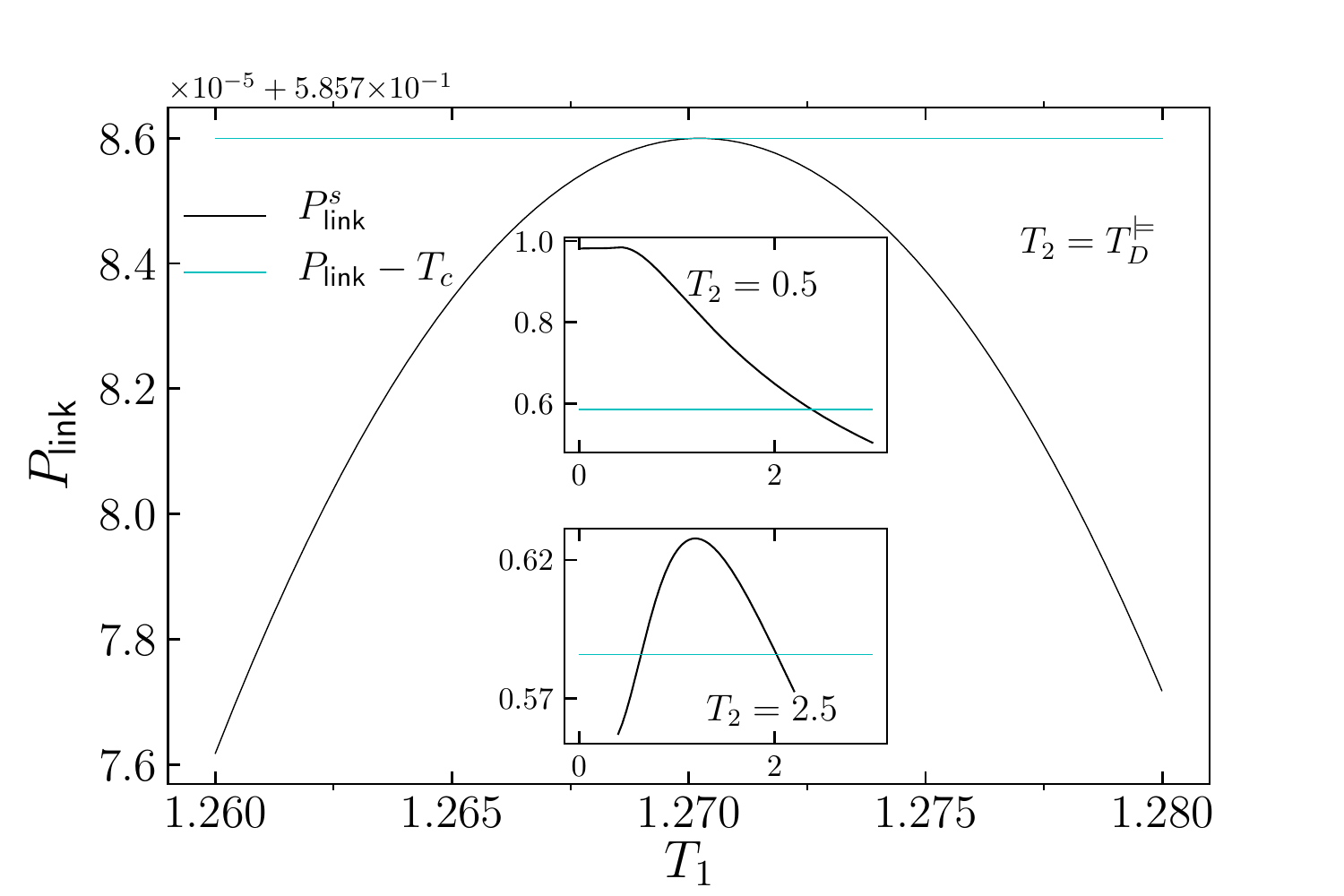}
		\caption{}
		\label{fig:lnk}
	\end{subfigure}
	\caption{(Color online):(a) Probability of adding a spin to the evolving cluster ($P_{\text{link}}$) in fixed $T_2 = T_D^{\models}$in term of $T_1$. Upper inset shows $P_{\text{link}}$  in term of $T_1$ for fixed $T_2 = 0.50$ and bottom inset shows ($P_{\text{link}}$)  in term of $T_1$ for fixed $T_2 = 2.50$. (b) Contour plot for Eq.(\ref{q8}) in $N=0$. The dash-line represent $1/\beta = T$ and markers show critical points that we find in simulation. }
	\label{p_link}
\end{figure*}

As be seen in Fig.(\ref{fig:md}) and its inset in $T_D^{\models}$ we only have one transition point (with the same exponents as other points, e.g. $\beta / \nu = 0.13 \pm 0.02$, and also $\gamma /\nu = 1.74 \pm 0.01$ as is seen in fig.~\ref{fig:Xid}) and system is unable to enter the ordered phase. When we go a bit above this point the system is always in the disordered phase as is shown in figs.~\ref{fig:dmd} and~\ref{fig:DXid}, especially in the inset of Fig.~\ref{fig:DXid} we observe that the exponent of $\chi_{max}$ for this case is $\mu = 1.62 \pm 0.02$. \\

As the final point, let us discus about how one can understand the form of the transition line, i.e. Fig.~\ref{phase} from an analytical point of view. To this end we compare the link probabilities of a single temperature Ising system (i.e. Eq.~\ref{q3}) with the one for two-temperature system (i.e. Eq.~\ref{q4}). The fact that all points in the critical line have the same properties as the Ising universality class, leads us to try finding an equivalent effective Ising system with an effective temperature $T_{\text{eff}}$, or equivalently $\beta_{\text{eff}}\equiv 1/T_{\text{eff}}$. Since the critical properties of the Ising model in the vicinity of the critical point is reflected in the properties of the FK clusters, which itself depends on the value of $P_{\text{link}}$, for the equivalent system we try the following equality, i.e. Eqs.~\ref{q3} and~\ref{q4}
\begin{align}
1-e^{-2\beta_{\text{eff}}} = \frac{1-e^{-2\beta_1}}{1+e^{2(\beta_1-\beta_2)}}+\frac{1-e^{-2\beta_2}}{1+e^{-2(\beta_1-\beta_2)}}.
\label{q7}
\end{align}
The real solution of this equation with respect to $\beta_{\text{eff}}$ is
\begin{align}
\beta_{\text{eff}}(\beta_1,\beta_2) = \frac{1}{2}\left[  \log(\frac{e^{2\beta_1} + e^{2\beta_2}}{e^{2(\beta_1 - \beta_2)}+ e^{2(\beta_2-\beta_1)}})\right].
\label{q8}
\end{align}
The contour plot of this solution is shown in Fig.~\ref{fig:plnk}, in which the critical points that we found in this paper (blue bold circles) coincide with the contour line corresponding to $\beta_{\text{eff}}(\beta_1,\beta_2) = T_c^{\text{Ising}}$. On this line, the corresponding FK cluster becomes critical (showing fractal properties~\cite{vasseur2012critical,janke2004geometrical}), undergoing a percolation transition, alongside with order-disorder transition. Our results above show that the effective system shows the properties of the Ising model, i.e. the same critical properties as the Ising universality class. \\

As a consistency check, we analyzed the crossing points of the left and right hand side of Eq.~\ref{q8} in fig.~\ref{fig:lnk}, where it is shown that for example when one sets $T_2=T_D^{\models}$ the curves cross each other at $T_1=T_D^\vdash$ as expected. The insets show two other situations. In upper inset of fig.~\ref{fig:lnk} for $T_2 =0.50 $ the crossing takes place in $T_1= 2.376 =T_1^{c_1}$, just where we found the critical point, and also for $T_2 =2.50 $ two crossing points are found. For all temperature  $T_2 >T_D^{\models} $ the two graphs never cross.\\

\section{conculation}
We as an example of superstatistic critical phenomena (SCP), we consider the Isning model with a distribution of temperature. This distribution was considered to be a binary one (with two temperatures $T_1$ and $T_2$) as the simplest generalization. This two-temperature Ising model was numerically simulated on square lattice with Monte Carlo method. We developed Metropolis and SW algorithms for this system using the analogy with one-temperature system, and also corresponding to two-level Boltzmann factor. We numerically showed that the system undergo an order-disorder transition which defines a critical line in $(T_1,T_2)$ phase space, see Fig.~\ref{phase}. The critical points were found using the data collapse analysis, as well as the Binder's cumulant method, which are consistent with the points that the heat capacity and the magnetic susceptibility show peak. For all temperature under $T_c^{\text{Ising}}$ (the critical temperature of the ordinary Ising system) one second-order phase transition was observed, whereas for $T_c^{\text{Ising}}<T_2<T_D^{\models}$ two second-order phase transition until was seen, and for $T_2>T_D^{\models}$ no transition takes place. Our numerical estimation of the critical exponents corresponding to the heat capacity and magnetic susceptibility, and also the exponents of the order parameter (average magnetism) all illustrate that all points on the critical line belong to the ordinary Ising universality class.  \\

To understand the structure of the critical line, we made an analogous effective system with a link probability $P_{\text{link}}(\beta_{\text{eff}})$ that is identical to the one for the binary temperature system, i.e. $P_{\text{link}}(\beta_1,\beta_2)$. Using this we obtained an analytical expression for the critical line, which maches perfectly with the numerical results, see fig.~\ref{fig:plnk}. This study can be generalized to other more sophisticated distribution of temperature, and also the systems where the temperature is distributed throughout the system in such a way that each cell takes one fixed temperature. These form our ideas for further studying the superstatistic critical phenomena.

\vspace{0cm}
\appendix

\section{ $n-\beta$ superposition}
\label{ap}
The Potts model is a generalization of the Ising model to more than two components (q-state). This system in addition to the theoretical aspect that it investigate critical properties in order-disorder phase transition~\cite{wu1982potts}, it is also possible to realize  the Potts model in experiments~\cite{PhysRevB.18.2209}. The Potts model is related to a number of other outstanding problem in lattice statistic like Vertex model~\cite{temperley1971relations}, percolation($q =1$ limit)~\cite{essam1980percolation}, resistor network($q =0 $limit)~\cite{fortuin1972random}. These are reasons to give motivate for investigate the Ising model in more general form and combination Potts model with superstatistic concept.  In more general form of  Ising model, in addition to $q-state$($q$ spin component) we add $n- temperature$ component with a same probability and propose $q$-state $n$-temperature Potts model. \\
Let us define The probability distribution of $\beta$ as follows:
\begin{equation}
f(\beta)=\frac{\delta (\beta - \beta_1)+\delta (\beta - \beta_2)+...+\delta (\beta - \beta_n)}{n},
\label{qq0}
\end{equation}
Using Eq.(\ref{bf}), a generalized $n$-level Boltzmann factor(nLBF) is obtained as follows:
\begin{equation}
B_n(E) = \frac{1}{n}(e^{-\beta_1 E}+e^{-\beta_2 E} +...+e^{-\beta_n E}),
\label{qq1}
\end{equation}
the changes in total energy due to the single spin flip is changed to  $E' = E + \delta E$. According to Eq.\ref{q10} for $p^{n\text{LBF}}(\beta_1,\beta_1,...,\beta_n)$

\begin{multline}
 p^{n\text{LBF}}(\beta_1,\beta_1,...,\beta_n)\equiv \frac{B_n(E')}{B_n(E)}=\\
 \frac{e^{-\beta_1 E'}+ e^{-\beta_2 E'}+...+e^{-\beta_n E'}}{e^{-\beta_1 E}+e^{-\beta_2 E}+..+e^{-\beta_n E}}.
\label{qq2}
\end{multline}
We simplify the above equation
\begin{multline}
 p^{n\text{LBF}}=\frac{e^{-\beta_1 \delta E}}{1+e^{-(\beta_2 - \beta_1)E}+e^{-(\beta_3 - \beta_1)E}+...+e^{-(\beta_n - \beta_1)E}} +\\ \frac{e^{-\beta_2 \delta E}}{1+e^{-(\beta_1 - \beta_2)E}+e^{-(\beta_3 - \beta_2)E}+...+e^{-(\beta_n - \beta_2)E}}+...+\\ \frac{e^{-\beta_n \delta E}}{1+e^{-(\beta_2 - \beta_n)E}+e^{-(\beta_3 - \beta_n)E}+...+e^{-(\beta_{n-1} - \beta_n)E}}
\label{qq3}.
\end{multline}
It is can be written in a series form as follows:
\begin{equation}
 p^{n\text{LBF}} =\sum_{i=1}^{n} \frac{e^{-\beta_i \delta E}}{\sum_{j=1}^{n} e^{-(\beta_j - \beta_i)E}}.
\label{qq4}
\end{equation}
At last in the SW algorithm:
\begin{equation}
P^{n}_{\text{link}} =\sum_{i=1}^{n} \frac{1- e^{-2 \beta_i}}{\sum_{j=1}^{n} e^{(\beta_j - \beta_i)}}.
\label{qq5}
\end{equation}
In this case we would have $n$-dimensional super-surface Ising universality class.
%----------------------------------------------------------------------------------------
%	BIBLIOGRAPHY
%----------------------------------------------------------------------------------------

\bibliography{refs}

%\bibliography{sample}

%----------------------------------------------------------------------------------------

\end{document}